\documentclass[aip,jcp,reprint,amsmath,amssymb,floatfix]{revtex4-1}
\usepackage[utf8]{inputenc}
\usepackage[T1]{fontenc}
\usepackage{mathptmx}
\usepackage{bm}
\usepackage{mathtools}
\usepackage{braket}
\usepackage{siunitx}
\usepackage{graphicx}
\usepackage{hyperref}

\usepackage{xspace}
\makeatletter
\def\ie{\textit{i.e.}\@\xspace}
\def\eg{\textit{e.g.}\@\xspace}
\makeatother
\newcommand{\abinitio}{\emph{ab~initio}\xspace}
\newcommand{\aposteriori}{\emph{a~posteriori}\xspace}

\renewcommand{\vec}[1]{\mathrm{\mathbf{#1}}}
\newcommand{\de}{\,{\mathrm{d}}}
\newcommand{\transp}{^\intercal}
\DeclareMathOperator\erfc{erfc}
\DeclareMathOperator{\Op}{{\hat O}}
\DeclareMathOperator{\OpS}{{\hat S}}
\DeclareMathOperator{\OpJ}{{\hat J}}

\DeclareMathOperator{\OpNO}{{\hat P_N^{\vec O}}}
\DeclareMathOperator{\OpNprime}{{\hat {P}^\prime_N}}
\DeclareMathOperator{\OpNU}{{\hat P_N^{\vec 1}}}

\graphicspath{{fig/}}

\usepackage[dvipsnames]{xcolor}
\newcommand{\changed}[1]{\textcolor{red}{#1}}
\renewcommand{\changed}[1]{#1}
\newcommand{\extref}[1]{#1}

\usepackage{fancyhdr}
\begin{document}

\title{Impact of quantum-chemical metrics\\
on the machine learning prediction of electron density}

\author{Ksenia R. Briling}
\affiliation{Laboratory for Computational Molecular Design, Institute of Chemical Sciences and Engineering,
\'{E}cole Polytechnique F\'{e}d\'{e}rale de Lausanne, 1015 Lausanne, Switzerland}
\author{Alberto Fabrizio}
\affiliation{Laboratory for Computational Molecular Design, Institute of Chemical Sciences and Engineering,
\'{E}cole Polytechnique F\'{e}d\'{e}rale de Lausanne, 1015 Lausanne, Switzerland}
\affiliation{National Centre for Computational Design and Discovery of Novel Materials (MARVEL),
\'{E}cole Polytechnique F\'{e}d\'{e}rale de Lausanne, 1015 Lausanne, Switzerland}
\author{Clemence Corminboeuf}
\email{clemence.corminboeuf@epfl.ch}
\affiliation{Laboratory for Computational Molecular Design, Institute of Chemical Sciences and Engineering,
\'{E}cole Polytechnique F\'{e}d\'{e}rale de Lausanne, 1015 Lausanne, Switzerland}
\affiliation{National Centre for Computational Design and Discovery of Novel Materials (MARVEL),
\'{E}cole Polytechnique F\'{e}d\'{e}rale de Lausanne, 1015 Lausanne, Switzerland}

\date{\today}

\begin{abstract}
Machine learning (ML) algorithms have undergone an explosive development impacting every aspect of computational chemistry.
To obtain reliable predictions, one needs to maintain the proper balance between the black-box nature
of ML frameworks and the physics of the target properties.
One of the most appealing quantum-chemical properties for regression models is the electron density,
and some of us recently proposed a transferable and scalable model
based on the decomposition of the density onto an atom-centered basis set.
The decomposition, as well as the training of the model,
is at its core a minimization of some loss function, which can be arbitrarily chosen
and may lead to results of different quality.
Well-studied in the context of density fitting (DF),
the impact of the metric on the performance of ML models has not been analyzed yet.
In this work, we compare predictions obtained using the overlap and the Coulomb-repulsion metrics for both decomposition and training.
As expected, the Coulomb metric used as both the DF and ML loss functions leads to the best results for the electrostatic potential and dipole moments.
The origin of this difference lies in the fact that the model is not constrained
to predict densities that integrate to the exact number of electrons $N$.
Since an \aposteriori correction for the number of electrons decreases the errors,
we proposed a modification of the model where $N$ is included directly into the kernel function,
which allowed to lower the errors on the test and out-of-sample sets.
\end{abstract}

\maketitle

\section{Introduction}

The molecular electron density $\rho(\vec{r})$
is one of the cornerstones of modern quantum chemistry and chemical physics.
Unlike the many-body wavefunction, the electron density, being a much simpler real-space scalar function,
is an observable and can be measured by X-ray diffraction\cite{KC2001} or transmission electron microscopy.\cite{MKPSKGCARISSK2011}
At the same time, as shown by the first Hohenberg--Kohn theorem,\cite{HK1964}
$\rho(\vec{r})$ embodies the same information as the wavefunction,
and thus gives access to all molecular properties
either directly or from its deformations in the presence of external fields.
Because of its fundamental role in electronic structure theory,
the electron density is a highly appealing target for machine learning (ML) models,
which is demonstrated by the growing number
of works on the non-linear regression of $\rho(\vec{r})$.
These models can be divided in two categories:
those treating the field as a set of values on a real-space grid\cite{ABXY2018,FPE2019,CKBKCR2019,JB2020}
and those built on a decomposition onto a basis set.\cite{BVLTBM2017}

In this second category, we have recently developed\cite{GFMWCC2019,FGMCC2019}
and demonstrated the wide-scope applicability\cite{FGMCC2019,FBGC2020a} and the generality\cite{FBGC2020b} of
a transferable model of the electron density. The model is based on symmetry-adapted
Gaussian process regression (SA-GPR)\cite{GWCC2018,github:TENSOAP}
and on a local decomposition of the electron
density field into an atom-centered spherical Gaussian basis. The decomposition,
as any density-fitting (DF) approximation, consists of two critical parts:
the selection\cite{FGMCC2019} or construction\cite{FBGC2020b} of a suitable basis set and
the determination of the basis set expansion coefficients.
The coefficients are determined by minimizing a loss function between the fitted
and the \abinitio densities. The set of density-decomposition coefficients
represent the target of the machine learning model.
During the training phase, the regression weights are found
by minimizing a second loss function, which reflects the
difference between the decomposed density and the predicted one.

In principle, any function of a set of real-space variables,
such as the electron density, can be exactly expanded
onto a complete set of basis functions in a unique way.
In practice, the auxiliary basis sets are incomplete
and the use of different loss functions leads
to different expansions of the electron density.

The simplest way to fit the approximate density to the original $\rho (\vec r)$
is to apply the least-squares technique\cite{N1969,BER1973,SF1975,V1988,VAF1993}
and find the decomposition coefficients $\{c^\mathrm{DF}_i\}$ that minimize the error
\begin{equation}
\label{eq:overlap}
\text{fitting error} = \int
\Big|\rho(\vec r)-\sum_i c^\mathrm{DF}_i\phi_i(\vec r)\Big|^2
\de^3 \vec r.
\end{equation}
This intuitive form of the fitting (decomposition) loss function can be re-stated
(usually under the constraint for the number of electrons)
in a more general quadratic functional of the density residue
$\Delta \rho(\vec r) = \rho(\vec r)-\sum_i c^\mathrm{DF}_i\phi_i(\vec r)$,
\begin{equation}
\label{eq:general}
\text{fitting error} = \iint
\Delta\rho(\vec r_1) \Op(\vec r_1, \vec r_2) \Delta\rho(\vec r_2) \de^3 \vec r_1 \de^3 \vec r_2,
\end{equation}
where $\Op(\vec r_1, \vec r_2)$ is a two-electron operator.
Eq.~\ref{eq:overlap} is a special case of Eq.~\ref{eq:general}
where $\Op$ is the overlap operator $\OpS = \delta(\vec r_{12})$.

The overlap fitting yields approximate densities that often lack accuracy.\cite{MD1982,VAF1993,SGHISW2000,JSGH2005}
For this reason, nowadays the standard procedure for density-fitting applications
is the electrostatic repulsion fitting\cite{W1973,DCS1979a,DCS1979b,HS1984,SH1986,ETOHA1995}
with $\Op$ being the Coulomb operator $\OpJ = |\vec r_{12}|^{-1}$,
which gives an approximate density whose electric field
is the closest to the original one.

The generality of Eq.~\ref{eq:general} promotes
other ways to find the decomposition coefficients.
For example, the anti-Coulomb metric\cite{GJPT1992} $\Op = -|\vec r_{12}|$,
although not widely used, gives an approximate density
with the closest electrostatic potential to the reference.
For extended systems, in order to avoid the slow decay of the Coulomb operator,
the complementary error-function Coulomb metric\cite{JSGH2005,RTKKHJHS2008}
\mbox{$\Op = \erfc(\omega|\vec r_{12}|)\cdot|\vec r_{12}|^{-1}$}
and the Gaussian-damped Coulomb metric\cite{RTKKHJHS2008}
\mbox{$\Op = \exp(-\omega|\vec r_{12}|^2)\cdot|\vec r_{12}|^{-1}$}
were also proposed.
Since both of them provide a smooth transition from the Coulomb ($\omega\to0$)
to the scaled overlap ($\omega\to\infty$) metric,
for our purposes it is sufficient to consider only the two limiting cases.

For the same reasons as the fitting of the \abinitio electron density,
the choice of the loss function to fix the regression weights is also not unique.
In fact, a simple regression model of the electron density
can also be formulated as a least-squares problem,
where the task is to find
the regression weights $\{x_j\}$ that minimize
a quadratic loss function
\begin{equation}
\label{eq:ml-overlap}
\Lambda(\vec x) =
\sum_{\substack{\rm training\\\rm set}}
\int\Big|\rho(\vec r) - \sum_i c^\mathrm{ML}_i(\vec x) \phi_i(\vec r)\Big|^2 \de^3 \vec r.
\end{equation}
Eq.~\ref{eq:ml-overlap} has the same structure
as the overlap density-fitting problem of Eq.~\ref{eq:overlap}
and can be generalized in the same fashion as Eq.~\ref{eq:general}.
The possibility to change the predicted expansion coefficients
simply by changing the metric both in the initial density decomposition
and in the regression loss function allows, in principle, the tuning of the SA-GPR machinery
for each specific application of the predicted electron density.

\changed{
In principle, it is possible to construct ML loss functions
using also other integrals targeted to DFT energies and energy densities,
\eg containing $\Delta\rho^{4/3}$ or reduced density gradients.
However, such loss functions cannot be written as quadratic functions of the regression weights,
and the learning step would require an iterative solution in a self-consistent manner.
}

Our previous works targeting the electron density with SA-GPR coincidentally exploited two different metrics
($\OpS$\cite{GFMWCC2019,FBGC2020b} and $\OpJ$\cite{FGMCC2019})
for the decomposition but only the overlap metric in
the machine learning loss function.
Given the known effects of the metric choice
in the density-fitting literature\cite{MD1982,VAF1993,SGHISW2000,JSGH2005}
and the lack of a corresponding systematic analysis for machine learning applications,
many questions remain unanswered.
For instance, is the density decomposed with one metric more difficult to learn than another?
Do the associated predicted densities differ significantly?
Which combinations of loss functions are the most efficient
for which application? More generally, these questions also address
a perhaps more fundamental
topic that is how do ML models interact with
deductive reasoning?

In the present work, we apply the
four possible combinations of $\OpS$ and $\OpJ$ metrics
on the same set of biologically-relevant molecules
and compare the quality of the predicted electron density
to reproduce different electronic properties, ranging
from the number of electrons to the dipole moments, electrostatic potentials (ESP), and
the characterization of the intra- and intermolecular electronic fingerprints with the density overlap
region indicator (DORI).\cite{SC2014}
As a result of this systematic analysis,
we also introduce several different
schemes to restore the correct number of electrons
in the predicted electron densities.

\section{Computational details}

This work uses
the side-chain--side-chain interaction subset of the BioFragment database\cite{BFZMSVMMS2017} (BFDb).
From the original set, we excluded molecules containing sulfur atoms and/or more than 25~atoms,
as well as several structures with unphysical atomic distances.
The final dataset contains 2287~dimers
and 35 of the most representative monomer structures.
Out of the total set, 2000~structures (1975~dimers and 25~monomers) were randomly selected for the training set
and 322~structures for the test set.

All quantum-chemical computations, except for three- and four-center overlap integrals,
were made with a locally modified version of PySCF.\cite{S2015,SBBBGLLMSSWC2017}
The reference density matrices were computed at the $\omega$B97X-D\cite{CH2008}/cc-pVQZ\cite{D1989} level
with the RI-JK approximation.
For density decomposition, the cc-pVQZ/JKFIT\cite{W2002} basis was used.

For sampling electrostatic potentials,
we computed molecular surfaces\cite{L2011}
$p(\vec r) = p_0$ with
\begin{equation}
p(\vec r) = \iint s(\vec r - \vec r_1) \rho_1(\vec r_1, \vec r_2) s(\vec r - \vec r_2) \de^3 \vec r_1 \de^3 \vec r_2,
\end{equation}
where $\rho_1(\vec r_1, \vec r_2)$ is the \abinitio one-particle density matrix,
$s(\vec r) = \exp(-a|\vec r|^2)$,
$a = 1/16$, and $p_0 = 1/16, 1/4, 1, 4$, or $32$.
The error in the predicted electrostatic potential $U_\mathrm{ML}(\vec r)$
with respect to the \abinitio one $U(\vec r)$ is defined as
\begin{equation}
\label{eq:esp-error}
\epsilon_\mathrm{ESP} =
\sqrt{\frac{ \iint_S (U(\vec r) - U_\mathrm{ML}(\vec r))^2 \de S}{ \iint_S \de S}},
\end{equation}
and the surfaces are discretized with the spherical quadrature rules.\cite{L1976,LL1999}

The density overlap region indicator\cite{SC2014}
was computed analytically on a cubic grid with a spacing of $0.1$~Bohr.
The comparison between two DORI fields in real space was done using the
Walker--Mezey similarity measure\cite{WM1994} $L(a,a')$
with $(a,a')=(0.1,0.7)$, $(0.7,0.95)$, and $(0.95,1)$.

The error in the predicted dipole moment $\boldsymbol \mu_\mathrm{ML}$
with respect to the \abinitio one $\boldsymbol \mu$ is defined as
\begin{equation}
\label{eq:dip-error}
\epsilon_\mathrm{dipole} =
| \boldsymbol \mu - \boldsymbol \mu_\mathrm{ML}|.
\end{equation}

For a density $\rho'(\vec r)$, we define the absolute
\begin{equation}
E_O[\rho'|\rho] = (\rho'-\rho|\Op|\rho'-\rho)
\end{equation}
and relative
\begin{equation}
e_O[\rho'|\rho] = E_O[\rho'|\rho] / (\rho|\Op|\rho)
\end{equation}
errors with respect to $\rho(\vec r)$
to be consistent with density-fitting and machine learning loss functions.

The tensorial $\lambda$-SOAP kernels\cite{GWCC2018,github:TENSOAP} were computed with
the following parameters:
environment cutoff $r_\mathrm{cut} = 4$~\AA,
Gaussian smearing $\sigma = 0.3$~\AA,
angular cutoff $l_\mathrm{cut} = 6$,
radial cutoff $n_\mathrm{cut} = 8$,
environmental kernel exponent $\zeta = 2$.
A subset of $M = 1000$ reference environments was taken
to reduce the dimensionality of the regression problem,
and the regularization parameter $\eta$ was set to~$10^{-6}$.

\section{The quantum-chemical metrics}
\label{sec:metric}
\subsection{Model construction}

Building a ML model for the electron density first consists in
fitting a linear combination of atom-centered basis functions $\{\phi_i\}$
\begin{equation}
\label{eq:rho-df}
\rho_\mathrm{DF}(\vec r) = \sum_{i} c^\mathrm{DF}_i \phi_i(\vec r)
\end{equation}
to the molecular electron density $\rho_\mathrm{QM}(\vec r)$,
which can be written in terms of the one-electron density matrix
or computed on a real-space grid etc.
The fitting coefficients $\{c^\mathrm{DF}_i\}$
are chosen to minimize a density-fitting (DF) loss function
\begin{equation}
\label{eq:df-lf}
\Lambda_\mathrm{DF}(\vec c^\mathrm{DF}) =
( \rho_\mathrm{DF} - \rho_\mathrm{QM} | \Op | \rho_\mathrm{DF} - \rho_\mathrm{QM}) \to \min,
\end{equation}
where $\Op$ is a two-electron operator
(overlap $\OpS = \delta(|\vec r_{12}|)$ or Coulomb repulsion $\OpJ = |\vec r_{12}|^{-1}$)
and the shorthand for two-electron integrals is
\begin{equation}
(f | \Op | g) =
\iint f(\vec r_1) \Op(\vec r_1, \vec r_2) g(\vec r_2) \de^3 \vec r_1 \de^3 \vec r_2.
\end{equation}
The solution for Eq.~\ref{eq:df-lf} is
\begin{equation}
\label{eq:df-solution}
\vec c^\mathrm{DF} = \vec O^{-1} \vec w,
\end{equation}
where
$ O_{ij} = (\phi_i | \Op | \phi_j) $ are the matrix elements of the operator $\Op$ and
$ w_i = (\phi_i|\Op|\rho_\mathrm{QM}) $ are, in the case of $O=S$,
the projections of the target field~$\rho_\mathrm{QM}$
onto the decomposition basis~$\{\phi_i\}$.
Different operators~$\Op$ yield different sets of coefficients~$\{c^\mathrm{DF}_i\}$,
each of which minimizes the loss function associated with~$\Op$.

In the same spirit, the ML loss function can be also written as
a sum over the structures of the training set (TrS)
\begin{equation}
\label{eq:ml-lf}
\Lambda_\mathrm{ML}(\vec x) =
\sum_\mathrm{TrS}
( \rho_\mathrm{ML} - \rho_\mathrm{DF} | \Op{}' | \rho_\mathrm{ML} - \rho_\mathrm{DF}) \to \min,
\end{equation}
where each ``predicted'' density
\begin{equation}
\label{eq:rho-ml}
\rho_\mathrm{ML}(\vec r) = \sum_{i} c^\mathrm{ML}_i(\vec x) \phi_i(\vec r)
\end{equation}
depends on the regression weights $\vec x$ via a kernel function
\begin{equation}
\label{eq:c=kx}
\vec c^\mathrm{ML}(\vec x) = \vec K \vec x,
\end{equation}
and $\Op{}'$ is also a two-particle operator.

The DF metric $O$ and the ML metric $O'$ are independent
and, in principle, can be chosen to be different.
For example, as we did in Ref.~\onlinecite{FGMCC2019},
it is perfectly possible to take $\Op = \OpJ$
so that the decomposed densities $\rho_\mathrm{DF}$
are the closest to the \abinitio densities $\rho_\mathrm{QM}$
in the sense that the self-repulsion of their residuals is the minimum,
and then take $\Op{}' = \OpS$ so that the
training-set predictions are (on average)
the closest to $\rho_\mathrm{DF}$ in the sense that
their overlap is the maximum.

However,
the use of different metrics for $O$ and $O'$ has a formally
unclear physical meaning.
On the other hand, using the same metric $O$ at both DF and ML steps
is analogous to the minimization of a loss function
\begin{equation}
\Lambda_\mathrm{DF+ML}(\vec x) =
\sum_\mathrm{TrS}
( \rho_\mathrm{ML} - \rho_\mathrm{QM} | \Op{} | \rho_\mathrm{ML} - \rho_\mathrm{QM}),
\end{equation}
making the predictions to be the closest to the original density in the $O$-sense,
as we did with the $S$-metric in Ref.~\onlinecite{GFMWCC2019}.

\subsection{Results}
\label{sec:initial_errors}

\begin{figure*}
\centering
\includegraphics[width=\linewidth]{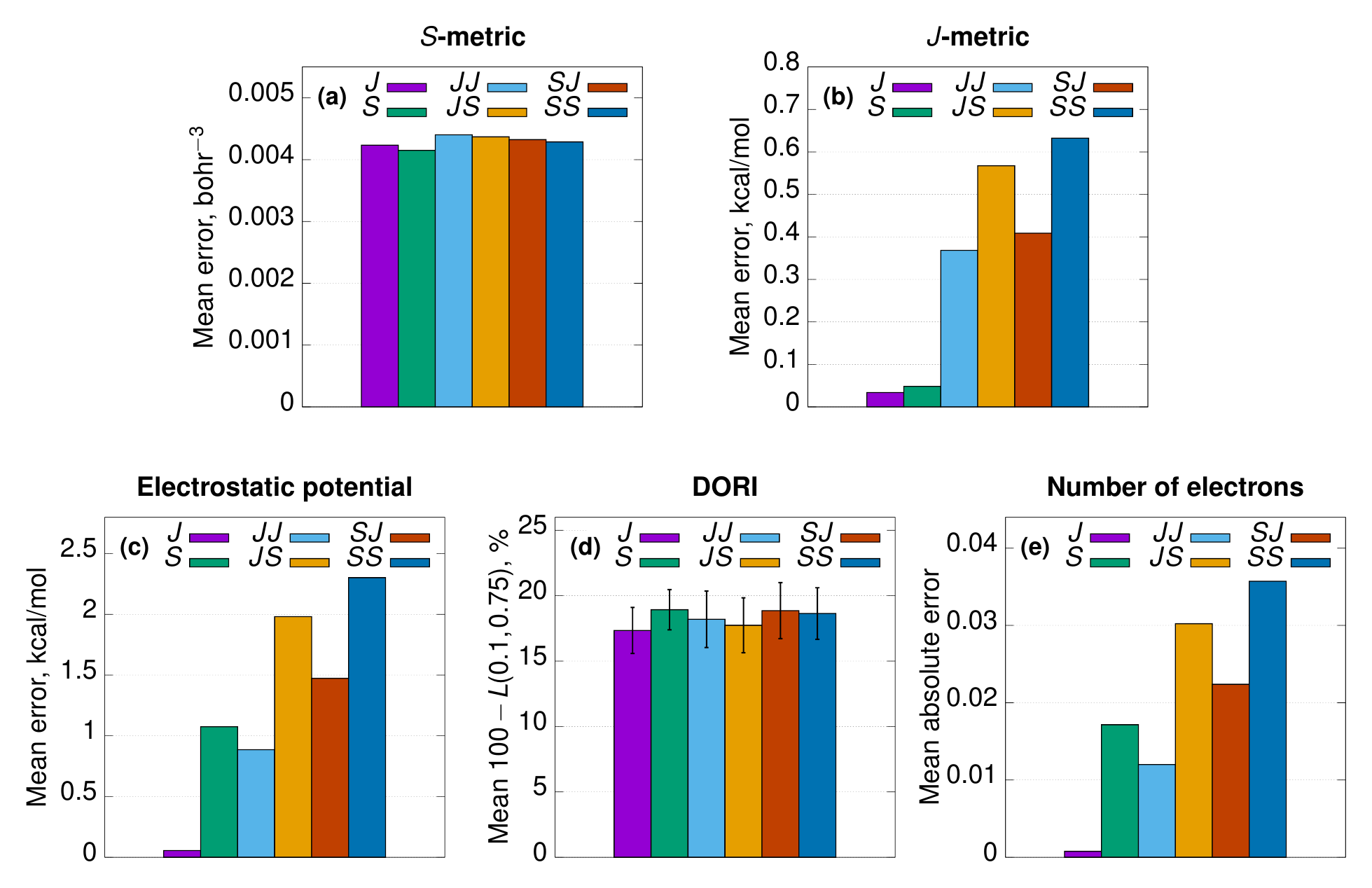}
\caption{
Mean error measures computed on the test set for fitted ($J$ and $S$) and predicted ($JJ$, $JS$, $SJ$, and $SS$) densities
without any constraints on the number of electrons:
\textit{(a)} $S$- and
\textit{(b)} $J$- metrics computed as $( \rho_\mathrm{ML} - \rho_\mathrm{QM} | \Op{} | \rho_\mathrm{ML} - \rho_\mathrm{QM})$;
\textit{(c)} errors in the electrostatic potential on the isosurface $p_0 = 4$,
             corresponding to average density $\braket{\braket{\rho}} \approx \SI{2e-6}{bohr^{-3}} $;
\textit{(d)} Walker--Mezey similarity indices for DORI in the non-covalent region (between 0.1 and 0.75),
             error bars show the standard deviation;
\textit{(e)} absolute errors in the number of electrons.
All errors are computed with respect to corresponding $\rho_\mathrm{QM}$.
\changed{Note that for the $S$-metric, we here use the $L^2$ and not the $L^1$ norm as in Ref.~\onlinecite{FGMCC2019} (see Fig.~S\extref{6} for comparison).}
}
\label{fig:initial_errors}
\end{figure*}

In this work,
four sets of densities were
predicted from the four possible combinations of DF and ML metrics
$OO'$: $JJ$, $JS$, $SJ$, and $SS$.
Comparison of the electron density mean errors $E_O[\rho|\rho_\mathrm{QM}]$
(\ie, with respect to $\rho_\mathrm{QM}$) for the two sets of fitted densities and four sets of predictions
can be found in Fig.~\ref{fig:initial_errors}a~and~\ref{fig:initial_errors}b.
Among the predictions, the lowest $J$-error is observed (on average) for the $JJ$-scheme,
where the $J$-metric is used for both DF and ML steps; the $SS$-scheme,
where the $J$-metric is not used at all, gives the highest $J$-error.
When using $S$-errors, the ranking is opposite.
It is not surprising, because
the goal of the framework is to yield the optimal predicted densities,
and what is optimal is defined by the DF and ML metrics.

However,
while the differences within the $S$-errors are less than 6\%,
the $J$-errors difference goes up to an order of magnitude.
It seems that the ML loss function with the $S$-metric has a more shallow minimum,
which can be already expected from the analysis of errors in the fitted densities alone
(see also Table~S\extref{4} in the Supplementary Material):
it is clear that the $J$-metric not only yields a smaller error in the number of electrons than the $S$-metric,
but is also more sensitive to small density differences.
It is interesting that the $JS$-scheme performs worse than the $SJ$, regardless of the error metric.
More detailed analysis shows
that the $S$- and $J$-errors are more sensitive to the DF and ML metric respectively.
This results in the error for $JS$ being larger than for $SJ$ in both cases.

The learning curves (Fig.~S\extref{1}a of the Supplementary Material),
which are based on the relative prediction errors (using the corresponding ML metrics)
with respect to their reference fitted densities,
show that the predictions are almost independent on the fitting metric
(the curves for the $JO$- and $SO$-schemes are nearly the same, for both $O=S$~or~$J$).
This trend essentially means that both the $S$- and $J$-fitting coefficients, $\{c^\mathrm{DF}_i\}$,
correlate with the atomic representation in a similar way.
It is known\cite{JSGH2005} that
the contribution $\sum_{j} (\vec O^{-1})_{ij}(\phi_j|\Op|\chi_p \chi_q) $
of an auxiliary function $\phi_i$, centered on one atom,
to the product of two basis functions $\chi_p$ and $\chi_q$, centered on another atom,
decays much slower with increasing distance between these two atoms
for repulsion metric than for the overlap one due to long-rangeness of the $\OpJ$ operator.
Potentially it can make $J$-coefficients harder to learn,
but in our set all the molecules were not big enough
to make the difference in decay noticeable.

On the other hand, the two sets of superimposed curves
($OS$ and $OJ$ on Fig.~S\extref{1}a of the Supplementary Material) do differ.
The fact that for the $OS$-curves the improvement from $250$ to~$2000$~training molecules
is slightly less significant than for the $OJ$-curves can be explained
by the lower sensitivity of the $S$-metric as discussed above.
Even though the $OS$-curves are lower than the $OJ$ ones,
it does not mean that the former predictions are in any way better:
these are pure prediction errors with respect to the corresponding fitted densities,
moreover, $S$-errors are shown for the $OS$-schemes and $J$-errors --- for the $OJ$-schemes.
In any case, the full-training-set prediction errors (the last points of the learning curves)
are very close (\num{5e-5}\% for $OS$ and \num{6e-5}\% for $OJ$) and the difference is not important.

\smallskip

To test these metric combinations on real-life applications,
we first chose two fundamentally different properties:
the electrostatic potential and the density overlap region indicator computed on all the test set molecules.
As shown in Fig.~\ref{fig:initial_errors}c~and~\ref{fig:initial_errors}d,
all four schemes lead to electrostatic potentials with a very different
quality with the $J$-metric consistently decreasing the error.
This result is not surprising considering that the $J$-metric yields an electron density
whose electric field is the closest to the reference one.\cite{DCS1979a,DCS1979b}
Because of the slow decay of the Coulomb potential,
the $J$-metric incorporates accurately long-range
information into the expansion coefficients
and it is generally preferred in
common quantum-chemical applications.\cite{MD1982,VAF1993,SGHISW2000,JSGH2005}

Since the model is not constrained to produce densities that integrate to the
correct number of electrons $N$ (see Sec.~\ref{sec:electrons:onemol}),
the errors $|\Delta N|$
are also shown on Fig.~\ref{fig:initial_errors}e.
The quality of the electrostatic potential does correlate with the error in the number of electrons
due to non-locality of both the properties.
On the other hand, DORI is much less sensitive to~$|\Delta N$|,
because it explicitly depends only on the local wave vector $\nabla\rho(\vec r)/\rho(\vec r)$.
As a simple example, a uniform scaling of a~reference density by a factor of $x$
leads to an error of $(x-1)\,N$ in the number of electrons
and thus to an error in the electrostatic potential, but with no influence on DORI.
Hence, the advantage of the $J$-metric observed above
arises from the fact that it usually yields a smaller error in the number of electrons.

Even for the test set, which is made of structures similar to those of the training set,
the predicted~$|\Delta N|$ can be as large as~$0.1$,
and even larger for out-of-sample molecules.
The error in the number of electrons leads to
the impossibility of reliably computing properties such as other multipole moments,
electrostatic potential, or exchange-correlation energy,
and thus prompts us to explore and compare different approaches to correct for~$N$
both after prediction and during the learning step.

\section{The number of electrons}
\label{sec:electrons}

In Sec.~\ref{sec:electrons:onemol}, we first discuss different ways to correct for the number of electrons
given by the approximate density of one molecule either at the decomposition or at the prediction step.
Next, in Sec.~\ref{sec:electrons:loss} and Sec.~\ref{sec:electrons:kernel},
we propose two modifications for the model
to exploit the information about the number of electrons at the learning step.

\subsection{\emph{A posteriori} correction of the predicted densities}
\label{sec:electrons:onemol}

By definition, the integral of the exact electron density over all space is the number of electrons $N$.
In our case, when an approximate density is determined by a set of coefficients~$\{c_i\}$, it integrates to a value
\begin{equation}
N(\vec c) = \int \rho(\vec r) \de^3\vec r = \sum_i c_i\,q_i,
\end{equation}
where
$q_i = \int \phi_i(\vec r) \de^3\vec r$ is the charge bearing by the basis function $\phi_i$.
Even though the loss function of Eq.~\ref{eq:df-lf}
searches for an approximate electron density being the closest to the reference,
it does not contain any explicit constraints,
and the fact that we use an incomplete basis set
leads to some inaccuracies in the number of electrons as in all other properties.
Moreover, the predicted coefficients in the form of Eq.~\ref{eq:c=kx}
are not constrained either and give a number of electrons close to $N$
only when the prediction errors are small enough.

The correct $N(\vec c)$ for the density fitting has been traditionally achieved
by adding a constraint\cite{SH1986} on the number of electrons
in the DF~loss function~\eqref{eq:df-lf}.
Hence, we get another set of decomposition coefficients
\begin{equation}
\label{eq:c-corr}
\vec c^{\mathrm{DF},N} = \vec c^\mathrm{DF} + \lambda \, \vec O^{-1}\vec q,
\end{equation}
where $\vec c^\mathrm{DF}$ is determined by Eq.~\ref{eq:df-solution}
and the Lagrange multiplier
$\lambda = (N-N(\vec c^\mathrm{DF}))/\vec q\transp\vec O^{-1}\vec q$.
Even though $q_i \neq 0$ only for spherically symmetric functions,
all the basis functions are coupled via the matrix $\vec O$
and thus participate in the correction of the coefficients.

It is also possible to rewrite Eq.~\ref{eq:c-corr} by introducing
a non-linear ``operator''~$\OpNO$ that acts on
any density in the form of a sum of atom-centered contributions (Eq.~\ref{eq:rho-df}~or~\ref{eq:rho-ml}),
giving a new density, which is the closest in $O$-sense to the ``old'' one, but integrates to exactly $N$ electrons,
\begin{equation}
\label{eq:refit}
\OpNO \vec c = \vec B\, \vec c + \vec n
\end{equation}
with
\begin{equation}
\label{eq:BB}
\vec B =  \vec1 - \frac{\vec O^{-1}(\vec q \vec q\transp) }{\vec q\transp \vec O^{-1}\vec q}, \quad
\vec n = \frac{N}{\vec q\transp \vec O^{-1}\vec q} \cdot\vec O^{-1}\vec q.
\end{equation}

In this way it is possible to correct (to ``refit'')
the coefficients $\vec c^\mathrm{ML}$ predicted by the original --- uncorrected --- machine learning model
and obtain the new coefficients \mbox{$\vec c^{\mathrm{ML},N} = \OpNO \vec c^\mathrm{ML}$}
suitable for computing electrostatic potential,
multipole moments, and other extensive properties.
Here it is implied that the correction metric $O$
is the same as the ML metric, but this restriction is not compulsory.

For instance, to correct the coefficients predicted for large molecules such as proteins,
the straightforward computation of the dot product $\vec O^{-1}\vec q$ ($O=S\text{ or }J$) is nearly impossible,
even though it can be implemented with integral screening and iterative matrix inversion methods.
Since we, in principle, can use different metrics for the decomposition, prediction,
and correction for the number of electrons,
a way to avoid the computational burden of inverting $\vec S$ or $\vec J$
is to simply use the unit matrix instead, \ie $\vec O = \vec 1$, and use an operator $\OpNU$.
\changed{(This operator only acts on the $s$-function coefficients.)}

Alternatively, an even simpler way to correct the coefficients for the number of electrons
is to scale them as
\begin{equation}
\label{eq:scale}
\OpNprime \vec c^\mathrm{ML} = \frac{N}{\vec q\transp \vec c^\mathrm{ML}} \cdot \vec c^\mathrm{ML}.
\end{equation}
This approach is somewhat arbitrary since
all the coefficients, even the ones that do not contribute to the number of electrons,
are scaled uniformly (one might as well scale only the coefficients for $s$-functions).
In this work, we use refitting with the unit matrix and scaling of the coefficients
only to correct the final predictions
in order to compare them with a more solid approach of Eq.~\ref{eq:refit}.

\subsection{Constrained learning: from $\bf M_0$ to $\bf M_L$}
\label{sec:electrons:loss}

In Sec.~\ref{sec:electrons:onemol}
we described how to correct for the number of electrons by
\aposteriori modification of the predictions obtained
from the original model $\bf M_0$.
These procedures are independent from the regression framework
and the number of electrons is never taken into account during the learning step.
However, such information could improve the final result.
Below, we consider two possibilities to explicitly
include the particle number information into the machine learning model.

In the original framework\cite{GFMWCC2019,FGMCC2019} $\bf M_0$,
the coefficients for a molecular system $m$ depend on
the regression weights $\vec x$ via the kernel matrix $\vec K_m$ (Eq.~\ref{eq:c=kx}),
and the working equations following from Eq.~\ref{eq:ml-lf}
(without regularization) are
\begin{equation}
\label{eq:x=B^-1A}
\vec x =
\big(\sum_{\mathclap{m\in\mathrm{TrS}}} \vec K_m\transp \vec O_m \vec K_m\big)^{-1}
\big(\sum_{\mathclap{m\in\mathrm{TrS}}} \vec K_m\transp \vec b_m\big)
\equiv \vec A^{-1} \vec u,
\end{equation}
where $\vec b_m = \vec O_m \vec c^\mathrm{DF}_m$ comes from density-fitting coefficients.
(If the DF and ML metrics are the same, $\vec b \equiv \vec w$.)

Using the Lagrange multipliers method, it is possible to
constrain the model to yield $N_m$~electrons
for each structure in the training set (or, generally, in the ``constraint set'' CS),
\begin{equation}
\begin{split}
\Lambda_\mathrm{ML}' &= \sum_{m\in\mathrm{TrS}}
(\rho'_{m,\mathrm{ML}}-\rho_{m,\mathrm{DF}}|\Op|\rho'_{m,\mathrm{ML}}-\rho_{m,\mathrm{DF}})
\\ &+ 2 \sum_{m\in\mathrm{CS}} \lambda_m \bigg(\int \rho'_{m,\mathrm{ML}} \de^3\vec r - N_m\bigg).
\end{split}
\end{equation}
The regression weights are
\begin{equation}
\label{eq:xprime}
\vec x' = \vec A^{-1} \vec u - \sum_{\mathclap{m\in\mathrm{CS}}} \vec K_m\transp \vec q_m \,\lambda_m
\end{equation}
and the Lagrange multipliers $\{\lambda_m\}$ are the solution of the linear system
\begin{equation}
\label{eq:ml-lambda}
\sum_{n\in\mathrm{CS}} (\vec q_m\transp \vec K_m \vec A^{-1} \vec K_n\transp \vec q_n) \lambda_n
= \vec q_m\transp \vec K_m \vec A^{-1} \vec u - N_m
\quad \forall m\in\mathrm{CS}.
\end{equation}
(We denote the constrained model $\bf M_L$ to distinguish it from the original $\bf M_0$.)

By construction, the regression weights $\vec x'$ lead to
coefficients giving the exact number of electrons
for any structure in the training set.
Yet, the predicted coefficients for an arbitrary molecule are not under any constraint
and should thus be corrected after prediction.
A smaller error in the number of electrons is however expected
in comparison to the one from the original model.

In principle, we are not restricted to put constraints on the same structures
as used for the minimization of $\Lambda_\mathrm{ML}'$.
The sums in Eqs.~\ref{eq:xprime}~and~\ref{eq:ml-lambda}
can be computed over \eg only a part of the training set,
the training set and some additional structures, or a completely different set of structures.
Despite having a vague physical sense, this flexibility can be exploited
for better understanding the model (see Sec.~\extref{II} of the Supplementary Material),
for example, by varying the training-set size with constant constraint subset or vice versa.

\subsection{Modification of kernels: from $\bf M_L$ to $\bf M_K$}
\label{sec:electrons:kernel}

With $\bf M_L$, the information about the number of electrons
is explicitly used in the model, but only for the training-set molecules.
Another more consistent possibility
is to modify directly the kernel function
to ensure the exact number of electrons for any set of coefficients obtained through it.
In this way,
the molecules in both the training and test sets are treated on the same footing,
while the training-set prediction error (\ie ML loss function)
is minimized for the corrected densities.

Combining Equations \ref{eq:ml-lf}, \ref{eq:c=kx}, and \ref{eq:refit},
we get a new model $\bf M_K$,
\begin{equation}
\label{eq:lambdaprimeprime}
\Lambda_\mathrm{ML}'' = \sum_{m\in\mathrm{mol}}
(\rho''_{m,\mathrm{ML}}-\rho_{m,\mathrm{DF}}|\Op|\rho''_{m,\mathrm{ML}}-\rho_{m,\mathrm{DF}}),
\end{equation}
where the $\rho''_{m,\mathrm{ML}}$ are determined by coefficients $\vec c^{\prime\prime\mathrm{ML}}_m$,
\begin{equation}
\label{eq:c=nkx}
\vec c^{\prime\prime\mathrm{ML}}_m(\vec x) = {\OpNO}_m \vec c^\mathrm{ML}_m(\vec x) = {\OpNO}_m \vec K_m \vec x.
\end{equation}
The working equations become
\begin{equation}
\label{eq:xprimeprime}
\vec x'' =
\big(\sum_{\mathclap{m\in\mathrm{mol}}} \vec K_m\transp \vec{\tilde O}_m \vec K_m\big)^{-1}
\big(\sum_{\mathclap{m\in\mathrm{mol}}} \vec K_m\transp \vec{\tilde b}_m\big)
\end{equation}
with $\vec {\tilde O} = \vec B\transp\vec O \vec B$ and $\vec {\tilde b} = \vec B\transp\vec b$
(since $\vec B\transp\vec n = 0$).
Equation~\ref{eq:xprimeprime} has the same form as Eq.~\ref{eq:x=B^-1A},
but the original quantum-chemical data $\vec {O}$ and $\vec {b}$
are transformed by matrix $\vec {B}$ defined by Eq.~\ref{eq:BB}.

The final predictions are obtained by using the regression weights $\vec x''$
in Eq.~\ref{eq:c=nkx},
which is analogous to the prediction according to Eq.~\ref{eq:c=kx} followed by the correction.

However, because the matrix $\vec B$ is idempotent, the modified metric matrix
$\vec {\tilde O} = \vec B\transp\vec O \vec B$
by construction has a zero eigenvalue and thus is singular,
making the regression problem ill-defined.
To get rid of the singularity, we propose to modify the loss function
and minimize the prediction error for both the corrected and uncorrected densities simultaneously,
by adding to Eq.~\ref{eq:lambdaprimeprime}
a small fraction $\alpha\in(0;1)$ of Eq.~\ref{eq:ml-lf},
\begin{equation}
\begin{split}
\Lambda_\mathrm{ML}'' =& \sum_{m\in\mathrm{mol}}  \Big(
\alpha\cdot (\rho_{m,\mathrm{ML}}-\rho_{m,\mathrm{DF}}|\Op|\rho_{m,\mathrm{ML}}-\rho_{m,\mathrm{DF}})
\\  &+
(1-\alpha)\cdot (\rho''_{m,\mathrm{ML}}-\rho_{m,\mathrm{DF}}|\Op| \rho''_{m,\mathrm{ML}}-\rho_{m,\mathrm{DF}})
\Big).
\end{split}
\end{equation}
The working equations are still in the form of~\eqref{eq:xprimeprime}
with the modified molecular data
\begin{gather}
\tilde{\vec O} = (1-\alpha)\cdot \vec B\transp\vec O \vec B + \alpha \cdot \vec O,\\
\tilde{\vec b} = (1-\alpha)\cdot\vec B\transp\vec b + \alpha\cdot\vec b,
\end{gather}
we used $\alpha=\num{e-6}$ to make the perturbation small
but still have an acceptable condition number of the $\tilde{\vec O}$ matrix.

\section{Metrics, models, and corrections: influence on ESP and dipole moment}
\label{sec:memo}

\begin{figure}[t]
\centering
\includegraphics[width=\linewidth]{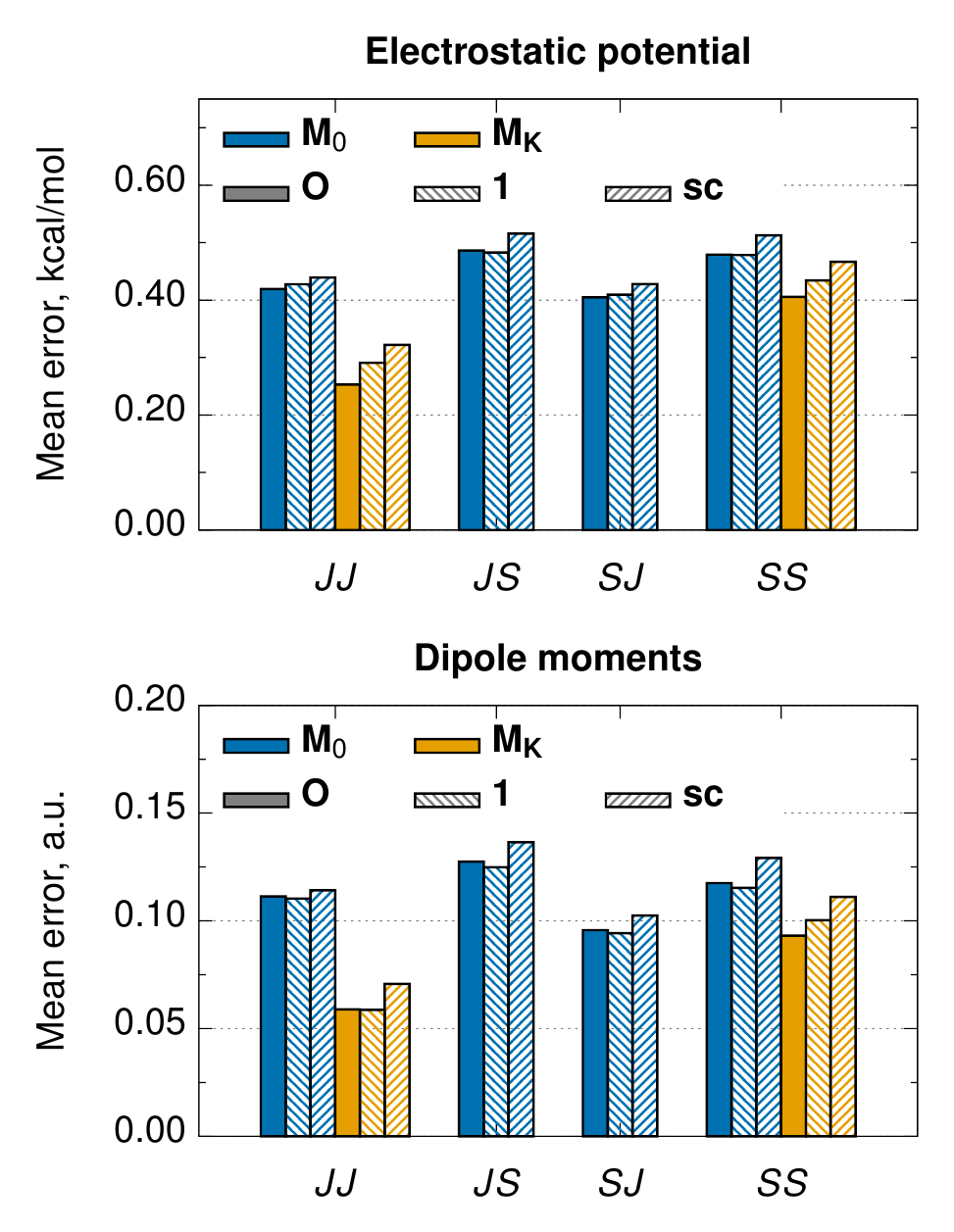}
\caption{
Comparison of four combinations of metrics ($JJ$, $JS$, $SJ$, and $SS$)
and two models ($\bf M_0$ and $\bf M_K$):
mean errors, computed on the test set,
in \textit{(top)} the electrostatic potential on the isosurface $p_0 = 4$
and \textit{(bottom)} dipole moments
for predicted densities upon correction for the number of electrons.
\textit{Solid bars} ($\vec O$): correction according to Eq.~\ref{eq:refit} with metric corresponding to the ML metric;
\textit{\textbackslash-filled bars} ($\vec 1$): correction using a unit matrix according to Eq.~\ref{eq:refit};
\textit{\slash-filled bars} (\textbf{sc}): correction by scaling according to Eq.~\ref{eq:scale}.
\changed{Errors in the ESP computed on other grids are provided in Table~S\extref{6}.}
}
\label{fig:esp_dipole}
\end{figure}

In addition to $\bf M_0$ (Sec.~\ref{sec:initial_errors}), we computed the predictions for
$\bf M_L$ (model of Eq.~\ref{eq:xprime}) and $\bf M_K$ (model of Eq.~\ref{eq:xprimeprime}).
However, even though $\bf M_L$ works as expected on a small set, the linear system of Eq.~\ref{eq:ml-lambda} becomes ill-defined
and the constraints cannot be fulfilled on a large enough training set
(number of molecules $\approx$ number of reference environments $M$, see Sec.~\extref{II} of the Supplementary Material for details).
For this reason, we have dropped $\bf M_L$ from the discussion and focus only on $\bf M_0$ and $\bf M_K$ hereinafter.

Figure~\ref{fig:esp_dipole} (solid bars) shows the errors with respect to the \abinitio results
for the electrostatic potentials and dipole moments predicted with $\bf M_K$ and $\bf M_0$
corrected according to Eq.~\ref{eq:refit}.
In comparison with the ESP predicted with the original $\bf M_0$ model (\ie, \mbox{0.8--2.3}~kcal/mol in Fig.~\ref{fig:initial_errors}c),
the prediction with an \aposteriori correction leads to errors at least two times smaller (about 0.4 kcal/mol).
In contrast and as expected, the DORI similarity measures are not affected by the correction
(See Fig.~S\extref{3} of the Supplementary Material).
Including the information about the number of particles into the kernel leads to lower errors in the ESP and dipole moments
than those with the \aposteriori correction alone for both the $JJ$ and $SS$ combinations.

We also explore the simpler ways to correct the final predictions, \ie,
the refitting with the unit matrix and the uniform scaling.
Even though the kernel function of $\bf M_K$ is already defined to always lead to the correct $N$, we also make,
for comparison, the final predictions with $\OpNU$ or $\OpNprime$ instead of $\OpNO$ in Eq.~\ref{eq:c=nkx}.
The errors in ESP and dipole moments are shown
on Fig.~\ref{fig:esp_dipole}, pattern-filled bars.
It is notable that correction with the unit metric does not significantly deteriorate the results obtained with the most
sophisticated scheme and can thus be used for larger molecules inadequate for $\OpNO$.

We note also that the effect of the metrics is more significant for the learning stage:
$S$-learning always gives larger errors than $J$-learning.
It is interesting that while $JJ$ and $SJ$ are nearly the same for ESP, $SJ$ works better for the dipole moments.
Yet, the $\bf M_K$ model in combination with the $JJ$ metric provides the best overall results on the test set.

\section{Extrapolation}

\begin{figure*}
\centering
\includegraphics[width=\linewidth]{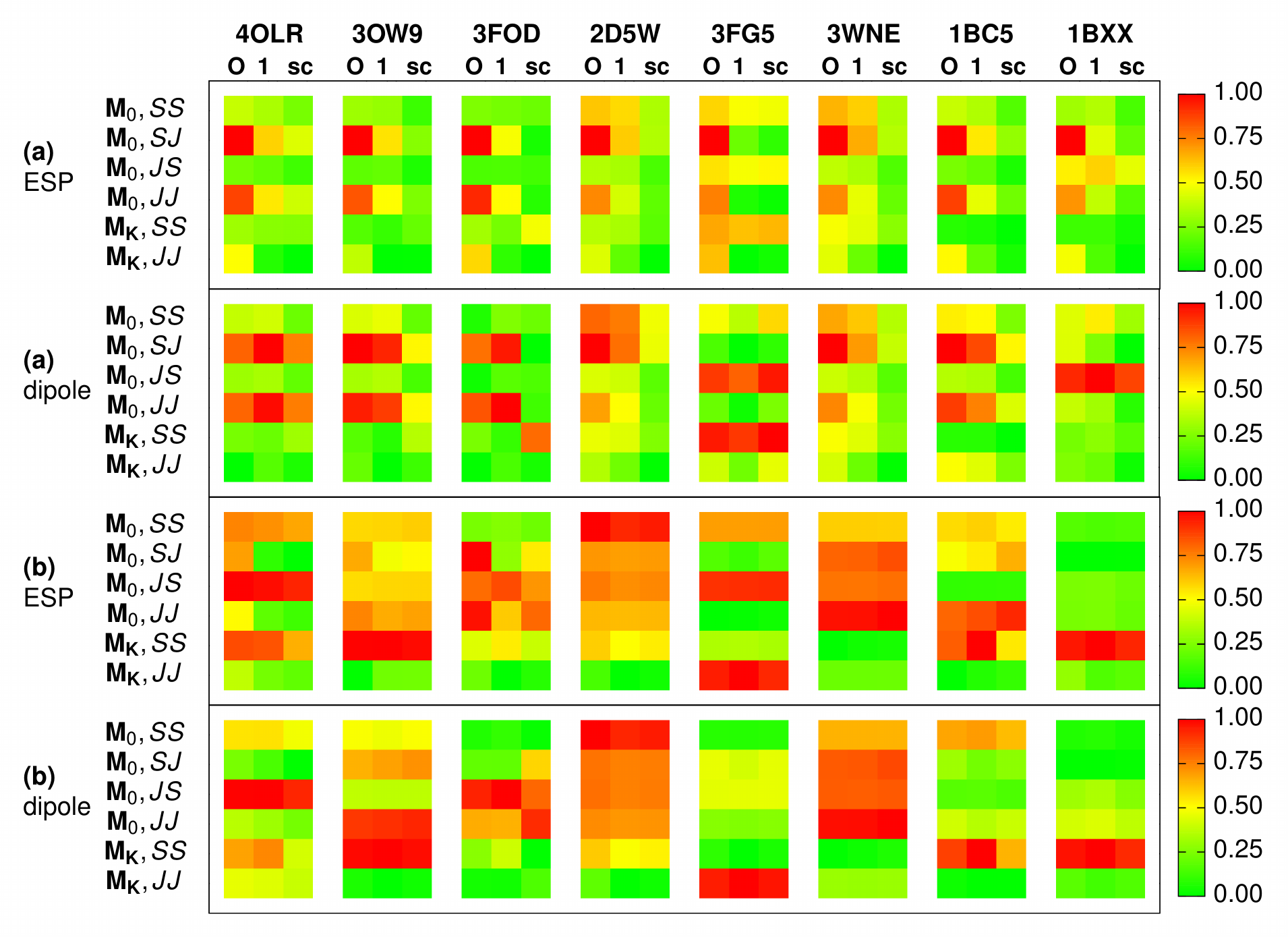}
\caption{
Comparison of the four combinations of metrics ($JJ$, $JS$, $SJ$, and $SS$),
two models ($\bf M_0$ and $\bf M_K$),
and three ways to correct for the number of electrons ($\vec O$, $\vec 1$, and \textbf{sc})
by prediction for eight oligopeptides (labeled by PDB~ID).
Each square represents an error in the predicted
electrostatic potential on the isosurface $p_0 = 0.125$ or
dipole moments of \textit{(a)} oligopeptides and \textit{(b)} ``no-backbone'' oligopeptides.
For the sake of clarity, the errors for each structure and property (\ie within each $3\times6$ rectangle)
are mapped to the $[0,1]$ interval,
\ie $\rm val \mapsto (val - val_{min}) / (val_{max} - val_{min})$.
}
\label{fig:oligo}
\end{figure*}

To validate the results of Sec.~\ref{sec:memo} on larger systems,
we predict the densities of the same eight oligopeptides taken from the Protein Data Bank
as used in our previous work\cite{FGMCC2019}
within both the $\bf M_0$ model with an \aposteriori correction and the $\bf M_K$ model.
Evaluating the performance of the corrected models on larger molecules is especially relevant
because the predictions of the electrostatic potential or multipole moments are not possible with the original $\bf M_0$ models.
The latter indeed yield to large errors in the number of electrons for these oligopeptides
(up to two orders of magnitude larger than those for the test set, see Table S\extref{9}),
which makes the computation of any property from the predicted density meaningless.

The errors in the predicted ESP and dipole moments
with respect to the \abinitio ones (normalized by mapping to the $[0,1]$~interval)
are shown on Fig.~\ref{fig:oligo}a.
To our surprise, the metrics $JJ$ and $SJ$ within $\bf M_0$,
which perform well on the test set, are usually the worst for the oligopeptides.
Moreover, the least physically sound correction scheme --- scaling ---
performs generally better than the sophisticated~$\OpNO$, suggesting an error cancellation.

Within this context, it is important to stress that our original training set is based only on the side-chain--side-chain
dimer subset of BFDb with no explicit representation of peptide bonds. For this reason, the highest absolute errors
in the predicted densities were shown to be mostly localized on the oligopeptide backbones.\cite{FGMCC2019}
In order to distinguish the effect of increasing the system size from the one originating from the lack of peptide backbones in the training set,
the peptide bonds were ``cut'' and the amino and carboxyl groups were replaced with hydrogen atoms.
Already within the non-corrected $\bf M_0$ model, the average errors in the number of electrons for these ``no-backbone'' systems
are an order of magnitude smaller than those for the original structures (see Table S\extref{9}).
For the corrected models,
the normalized errors in the predicted properties are shown on Fig.~\ref{fig:oligo}b with the absolute errors shown in Fig.~S\extref{4}b.
On average, the absolute errors are \mbox{3--6}~times lower than those for the original oligopeptides,
which confirms the significant perturbation associated with the peptide bonds, while comparing the different models and corrections.
This problem, which is not the topic of this work, could be easily addressed by extending the training set.
The error spread (Fig.~S\extref{4}b) also decreases,
\textit{e.g.}, the ESP errors for 3OW9 lie between~$1.5$~and~$6.6$~kcal/mol,
whereas for its no-backbone version the interval is $(0.6, 1.6)$~kcal/mol.
Overall, all the models and metrics perform very similarly and lead to fairly impressive predictions.
Akin to the test set, the $\bf M_K$, $JJ$ combination with any correction scheme offers the best compromise
as it leads to the most accurate predictions for most oligopeptides.
Similarly, the performance of the \aposteriori corrected $\bf M_0$, $JS$ models,
which was slightly inferior for the test set, is also less robust for the oligopeptide set
\changed{(for additional comparisons on the oligopeptide set, refer to Figure~S\extref{5} and Table~S\extref{9} in the Supplementary Material)}.

\section{Conclusions}

The analysis of the interplay between deductive reasoning based on quantum-chemical knowledge
and the inductive nature of statistical learning is a fundamental issue
to further improve quantum machine learning models.
In this work, we analyze the effects of varying the quantum-chemical metrics
used for the decomposition and regression of the molecular electron density.
We find that the machine learning loss function is more affected by the choice of metric than the loss function associated with decomposition
but overall, the $JJ$-scheme shows the best performance.
Yet, the learning exercise is equally difficult regardless of the metric used to decompose the density.

Importantly, imposing the correct number of electrons appears crucial
to accurately predict extensive properties such as the ESP and multipole moments.
Correcting the predictions \aposteriori for the number of electrons makes the accuracy of the $\bf M_0$ model
largely independent from the choice of the quantum-chemical metric.
This result is especially important for periodic systems or for situations where the charge density (or another density-like object)
can be obtained only on a real-space grid for which the Coulomb metric is ill-defined and the overlap has to be used.
As a step forward, we propose the $\bf M_K$ model,
in which the kernels explicitly include the information about the number of electrons.
While both \aposteriori correction and kernel modification increase slightly
the computational complexity on the prediction step,
it is always possible to apply other corrections
(such as the \changed{unit-matrix} correction) when extrapolating on larger chemical systems.

Overall, this work demonstrates that choosing a proper quantum-chemical \changed{metric} to optimize ML models is important
and that this is especially true if the model is not built to encode all the proper fundamental physical constraints.
\footnote{\changed{While the framework can, in principle, accommodate any molecular property as constraint,
the computational advantages of machine learning are leveraged only when using readily obtainable quantities such as the number of electrons,
which do not require any quantum-chemical computation.}}

\section*{Supplementary material}
See the Supplementary Material for the learning curves,
discussion of the $\bf M_L$ model, additional numerical data, and statistical analysis.

\begin{acknowledgments}
The authors thank Andrea Grisafi, David M. Wilkins, and Michele Ceriotti
for sharing the code\cite{github:TENSOAP} to construct the tensorial SOAP kernels.
A.F. acknowledges financial support from the National Centre of Competence in Research (NCCR)
``Materials' Revolution: Computational Design and Discovery of Novel Materials (MARVEL)''
of the Swiss National Science Foundation (SNSF).
K.B. was supported by the European Research Council (ERC, grant agreement no 817977).
\end{acknowledgments}

\section*{Data availability}
The data and the model that support the findings of this study
are freely available on the Materials Cloud at
\href{https://doi.org/10.24435/materialscloud:d8-0h}{\url{https://doi.org/10.24435/materialscloud:d8-0h}}.

\bibliography{sj.bib}
\clearpage

\end{document}


\title{{\sc Supplementary Material}\texorpdfstring{\\}{}
Impact of quantum-chemical metrics\texorpdfstring{\\}{}
on the machine learning prediction of electron density}

\author{Ksenia R. Briling}
\affiliation{Laboratory for Computational Molecular Design, Institute of Chemical Sciences and Engineering,
\'{E}cole Polytechnique F\'{e}d\'{e}rale de Lausanne, 1015 Lausanne, Switzerland}
\author{Alberto Fabrizio}
\affiliation{Laboratory for Computational Molecular Design, Institute of Chemical Sciences and Engineering,
\'{E}cole Polytechnique F\'{e}d\'{e}rale de Lausanne, 1015 Lausanne, Switzerland}
\affiliation{National Centre for Computational Design and Discovery of Novel Materials (MARVEL),
\'{E}cole Polytechnique F\'{e}d\'{e}rale de Lausanne, 1015 Lausanne, Switzerland}
\author{Clemence Corminboeuf}
\email{clemence.corminboeuf@epfl.ch}
\affiliation{Laboratory for Computational Molecular Design, Institute of Chemical Sciences and Engineering,
\'{E}cole Polytechnique F\'{e}d\'{e}rale de Lausanne, 1015 Lausanne, Switzerland}
\affiliation{National Centre for Computational Design and Discovery of Novel Materials (MARVEL),
\'{E}cole Polytechnique F\'{e}d\'{e}rale de Lausanne, 1015 Lausanne, Switzerland}

\date{\today}

\maketitle
\onecolumngrid
\tableofcontents

\newpage
\section{Learning curves}

\begin{figure*}[h]
\centering
\includegraphics[width=0.91 \textwidth]{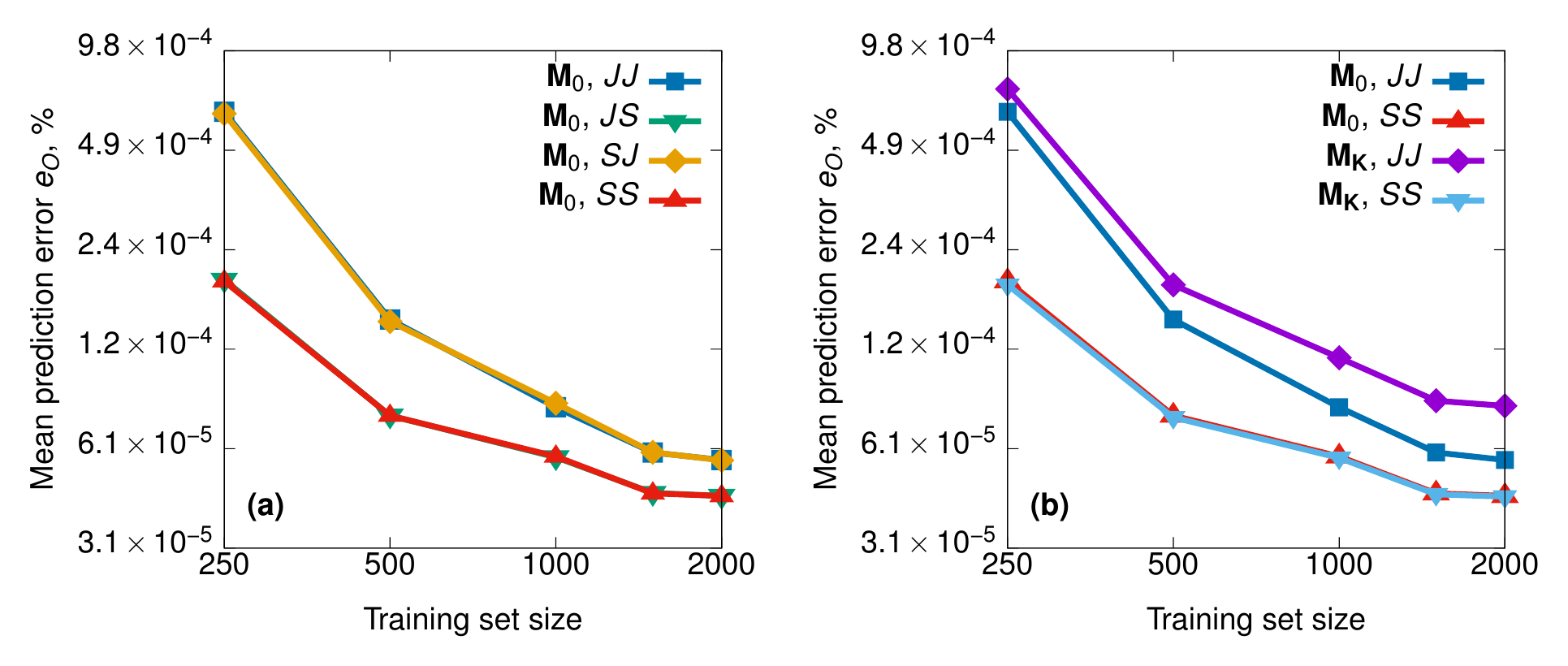}
\caption{
Comparison of the learning curves for \textit{(a,b)} four metric combinations within the $\bf M_0$ model;
\textit{(c,d)} two metric combinations and two models ($\bf M_0$ and $\bf M_K$).
Metric $O$ used to compute the errors corresponds to the ML (second) metric,
\ie, the $J$-metric for $JJ$ and $SJ$ and the $S$-metric for $JS$ and~$SS$.
The number of reference environments $M=1000$.\\
\textit{(a)}: the $JJ$ (blue) and $SJ$ (yellow) curves are superimposed, and $SS$ (red) and $JS$ (green) are superimposed.\\
\textit{(b)}: the ($\bf M_0$, $SS$) (red) and ($\bf M_K$, $SS$) (light-blue) curves are superimposed.
}
\label{fig:lc}
\end{figure*}

The learning curves are shown on Fig.~\ref{fig:lc} and the numerical data are provided in Table~\ref{tab:lc}.

The relative prediction errors $e_O$ are computed as
\begin{equation}
e_O
= \frac{(\rho_\mathrm{ML}-\rho_\mathrm{DF}|\Op|\rho_\mathrm{ML}-\rho_\mathrm{DF})}{(\rho_\mathrm{DF}|\Op|\rho_\mathrm{DF})}
=\frac{(\vec c_\mathrm{ML} - \vec c_\mathrm{DF})\transp \vec O (\vec c_\mathrm{ML} - \vec c_\mathrm{DF})}
{\vec c_\mathrm{DF}\transp \,\vec O\, \vec c_\mathrm{DF}^{\vphantom{\intercal}}}
\end{equation}
for $\bf M_0$
and as
\begin{equation}
e_O = \frac{(\vec c_\mathrm{ML} - \vec{\tilde c}_\mathrm{DF})\transp\vec{\tilde O}(\vec c_\mathrm{ML}-\vec{\tilde c}_\mathrm{DF})}
{\vec{\tilde c}_\mathrm{DF}\transp \,\vec{\tilde O}\, \vec{\tilde c}_\mathrm{DF}^{\vphantom{\intercal}}}
\end{equation}
for $\bf M_K$, where $\vec{\tilde c}_\mathrm{DF} = \vec{\tilde O}^{-1} \vec{\tilde b} \approx \vec c_\mathrm{DF}$,
while $\vec c_\mathrm{ML} = \vec K \vec x$ for both models.

\begin{table}[h]
\caption{\label{tab:lc}{Numerical data for Fig.~\ref{fig:lc}. }}
\begin{ruledtabular}
\begin{tabular}{cccccccc}
\multirow{2}{*}{Training set fraction}
       &  \multicolumn{3}{c}{$e_J$, \%}                            & \multicolumn{3}{c}{$e_S$, \%}  \\\cline{2-4}\cline{5-7}
       &  $\bf M_0$, $JJ$  &  $\bf M_0$, $SJ$  &  $\bf M_K$, $JJ$  &  $\bf M_0$, $SS$  &  $\bf M_0$, $JS$  &  $\bf M_K$, $SS$   \\ \hline
0.125  &  \num{6.38e-04}   &  \num{6.30e-04}   &  \num{7.48e-04}   &  \num{1.96e-04}   &  \num{1.98e-04}   &  \num{1.91e-04}    \\
0.250  &  \num{1.50e-04}   &  \num{1.48e-04}   &  \num{1.91e-04}   &  \num{7.68e-05}   &  \num{7.68e-05}   &  \num{7.59e-05}    \\
0.500  &  \num{8.14e-05}   &  \num{8.37e-05}   &  \num{1.15e-04}   &  \num{5.78e-05}   &  \num{5.75e-05}   &  \num{5.72e-05}    \\
0.750  &  \num{5.94e-05}   &  \num{5.95e-05}   &  \num{8.52e-05}   &  \num{4.47e-05}   &  \num{4.48e-05}   &  \num{4.44e-05}    \\
1.000  &  \num{5.64e-05}   &  \num{5.63e-05}   &  \num{8.22e-05}   &  \num{4.39e-05}   &  \num{4.39e-05}   &  \num{4.37e-05}    \\
\end{tabular}
\end{ruledtabular}
\end{table}

\clearpage
\section{Breakdown of \texorpdfstring{$\bf M_L$}{M\_L}}

In the initial tests,
even though the $\bf M_L$ model worked as expected on a $1/8$ (or $1/4$) of the training set
yielding the exact number of electrons for the molecules in the training subset used,
the constraints could not be fulfilled on the full training set (2000~structures)
because the equation on Lagrange multipliers $\{\lambda_n\}$
\begin{equation}
\label{eq:ml-lambda}
\sum_{n\in\mathrm{CS}} (\vec q_m\transp \vec K_m \vec A^{-1} \vec K_n\transp \vec q_n) \lambda_n
=\sum_{n\in\mathrm{CS}} L_{mn} \lambda_n
= \vec q_m\transp \vec K_m \vec A^{-1} \vec u - N_m
\quad \forall m\in\mathrm{CS}
\end{equation}
was ill-defined.

\begin{figure*}[p]
\centering
\includegraphics[width=0.7 \textwidth]{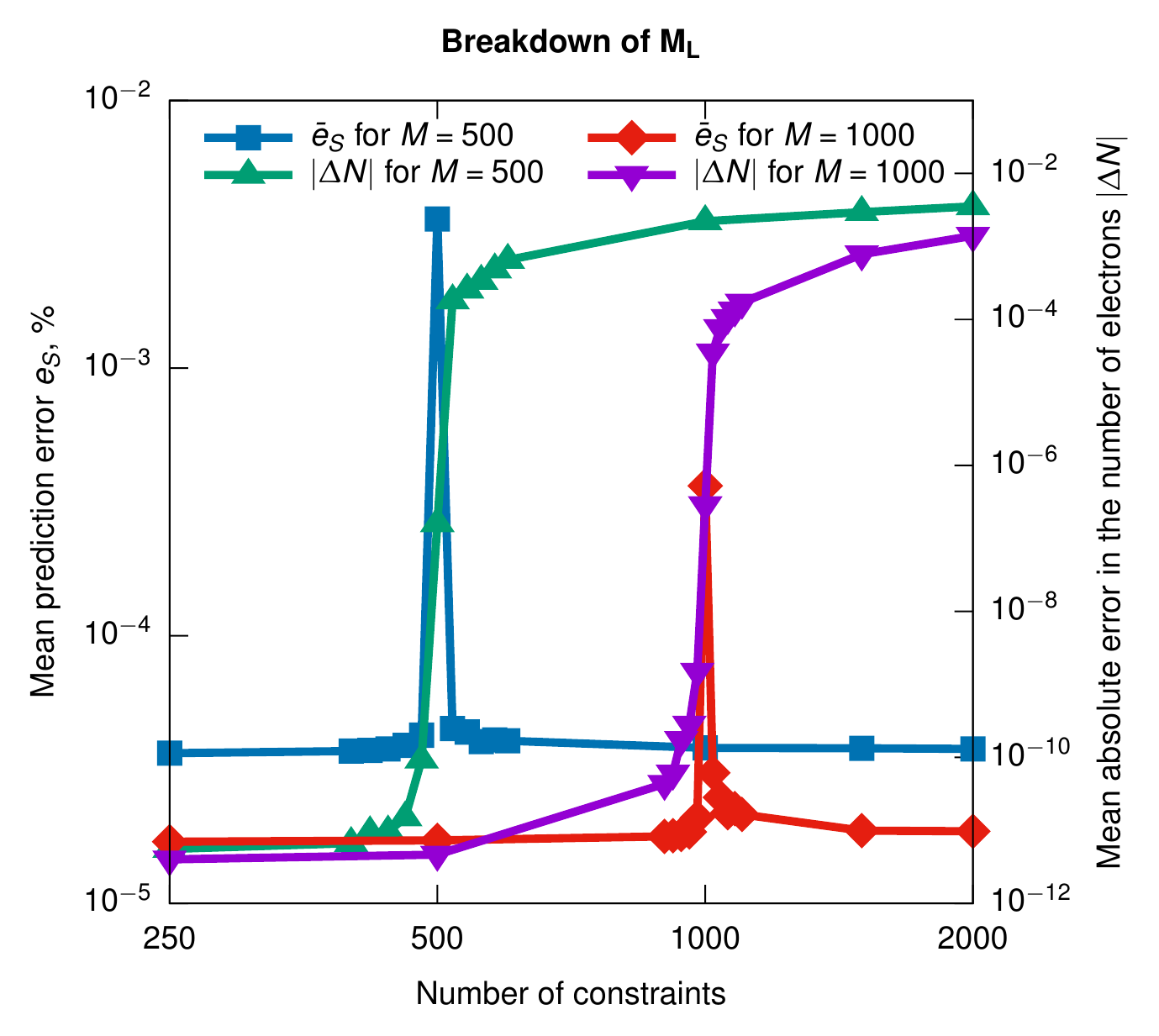}
\caption{
The prediction errors $e_S$ and errors in the number of electrons $|\Delta N|$ on the training subset
for different constraints subset sizes within the ($\bf M_L$, $SS$) model.
The size of the training subset is 250~molecules.
For both $M$ tested,
the error in the number of electrons $|\Delta N|$ dramatically increases and becomes much
larger than round-off errors when the number of constraints approaches $M$, indicating that the constraints are not fulfilled anymore.
The same is observed (but not shown) with the condition number of the $\vec L$ matrix.
}
\label{fig:ML}
\end{figure*}

To further analyze this behavior, we modified the model such as to minimize the ML loss function only on $1/8$ of the training set (250~molecules)
with constraints put on the molecules from a wider subset of the full training set (250--2000~molecules),
\begin{equation}
\{\text{training subset TrS}\} \subseteq \{\text{constraints subset CS}\} \subseteq \{\text{full training set}\},
\end{equation}
and computed the prediction errors $\bar e_O$ and the errors in the number of electrons $|\Delta N|$
on the training subset
for different constraints subset sizes (Fig.~\ref{fig:ML}).
Since Eq.~\ref{eq:ml-lambda} is ill-defined for some constraints subsets,
instead of the exact solution we took the least-squares one.
The breakdown of $\bf M_L$ is indicated by a dramatic increase of both $\bar e_O$ and $|\Delta N|$.
These results are qualitatively similar to
those obtained when the training subset and the constraints subset are the same,
but the former are much faster to obtain.

Due to a large ($10^4$--$10^5$) number of atoms in the training set,
a representative subset of $M$ ($\sim 10^3$) atoms (also called reference environments)
is usually chosen to reduce the dimensionality of the regression problem
\begin{equation}
\label{eq:x=B^-1A}
\vec x =
\big(\sum_{\mathclap{m\in\mathrm{TrS}}} \vec K_m\transp \vec O_m \vec K_m\big)^{-1}
\big(\sum_{\mathclap{m\in\mathrm{TrS}}} \vec K_m\transp \vec b_m\big)
\equiv \vec A^{-1} \vec u.
\end{equation}
Each element of $\vec x$ of Eq.~\ref{eq:x=B^-1A} represents a basis function
corresponding to one of the reference environments.
In this work, we first took $M=1000$ leading to \num{68161}~variables,
and then $M=500$ (\num{34685}~variables).
In both cases, the breakdown of $\bf M_L$ occurs when the constraints subset size approaches
the number of reference environments (Fig.~\ref{fig:ML}).

The matrix, which needs to be inverted to find $\{\lambda_n\}$ can be written as
\begin{equation}
\vec H = \vec V \transp \vec A^{-1} \vec V,
\end{equation}
where $\vec v_n = \vec K_n\transp \vec q_n$ and $\vec V$ is a matrix made of $\vec v_n$ columns.
Since
\begin{equation}
\rank(\vec M_1\vec M_2) \leq \min (\rank \vec M_1, \rank \vec M_2)
\end{equation}
and the matrix $\vec A^{-1}$ exists and thus has a full rank,
\begin{equation}
\rank \vec H \leq \rank \vec V.
\end{equation}
The elements of $\vec q_m$ are nonzero only for $s$-functions,
and all the $s$-functions centered on the same atom
share the same rows in $\vec K_m$. Thus the rank of $\vec V$ cannot be larger than $M$,
and the breakdown is inevitable.

\begin{table}[p]
\caption{\label{tab:ML}{Numerical data for Fig.~\ref{fig:ML}. }}
\begin{ruledtabular}
\begin{tabular}{cccccc}
\multicolumn{3}{c}{$M=500$}                                & \multicolumn{3}{c}{$M=1000$}            \\
Number of constraints & $e_S$, \%        & $|\Delta N|$    &  Number of constraints & $e_S$, \%      & $|\Delta N|$  \\ \hline
250                   & \num{3.64e-05}   & \num{5.6E-12}   &  250                   & \num{1.70e-05} & \num{4.0E-12} \\
400                   & \num{3.71e-05}   & \num{6.7E-12}   &  500                   & \num{1.72e-05} & \num{4.7E-12} \\
420                   & \num{3.73e-05}   & \num{9.4E-12}   &  900                   & \num{1.78e-05} & \num{4.4E-11} \\
440                   & \num{3.78e-05}   & \num{9.6E-12}   &  920                   & \num{1.79e-05} & \num{6.0E-11} \\
460                   & \num{3.89e-05}   & \num{1.5E-11}   &  940                   & \num{1.81e-05} & \num{1.7E-10} \\
480                   & \num{4.25e-05}   & \num{9.3E-11}   &  960                   & \num{1.85e-05} & \num{2.8E-10} \\
500                   & \num{3.61e-03}   & \num{1.6E-07}   &  980                   & \num{2.09e-05} & \num{1.5E-09} \\
520                   & \num{4.50e-05}   & \num{1.8E-04}   &  1000                  & \num{3.63e-04} & \num{2.9E-07} \\
540                   & \num{4.37e-05}   & \num{2.6E-04}   &  1020                  & \num{3.08e-05} & \num{3.6E-05} \\
560                   & \num{4.02e-05}   & \num{3.4E-04}   &  1040                  & \num{2.49e-05} & \num{7.6E-05} \\
580                   & \num{4.08e-05}   & \num{4.8E-04}   &  1060                  & \num{2.18e-05} & \num{1.1E-04} \\
600                   & \num{4.05e-05}   & \num{6.6E-04}   &  1080                  & \num{2.26e-05} & \num{1.3E-04} \\
1000                  & \num{3.81e-05}   & \num{2.2E-03}   &  1100                  & \num{2.15e-05} & \num{1.7E-04} \\
1500                  & \num{3.80e-05}   & \num{3.0E-03}   &  1500                  & \num{1.87e-05} & \num{7.8E-04} \\
2000                  & \num{3.78e-05}   & \num{3.5E-03}   &  2000                  & \num{1.86e-05} & \num{1.4E-03} \\
\end{tabular}
\end{ruledtabular}
\end{table}

\clearpage
\section{DORI, other numerical results, and statistical analysis}

\begin{figure*}[h]
\centering
\includegraphics[width=0.33\textwidth]{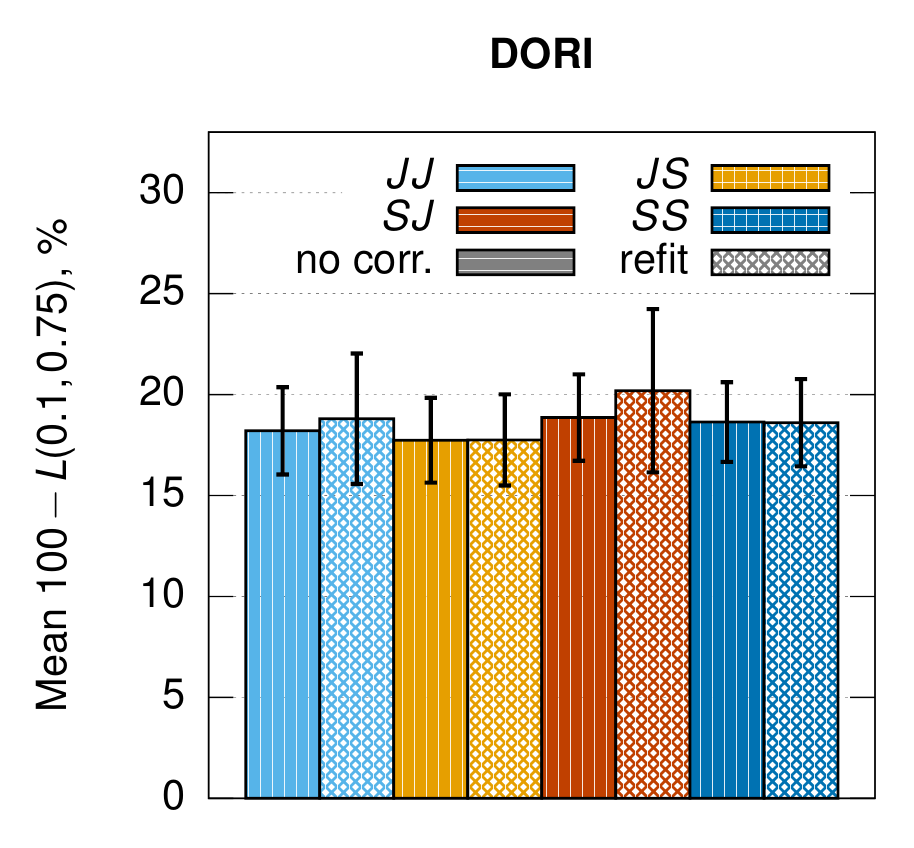}
\caption{
Mean Walker--Mezey similarity indices for DORI in the non-covalent region
computed on the test set for densities predicted within the $\bf M_0$ model
\textit{(solid bars)} without any constraints on the number of electrons
and \textit{(pattern-filled bars)} after the \aposteriori refitting correction;
error bars show the standard deviation.
}
\label{fig:dori}
\end{figure*}

\begin{table}[h]
\caption{\label{tab:initialerrors}{
\textbf{Numerical data for Fig.~1 and Fig.~\ref{fig:dori}}:
Density-fitting ($J$, $S$) and prediction ($JJ$, $JS$, $SJ$, $SS$) mean error measures
on the test set (322~molecules) without any corrections for the number of electrons
\textit{($S$-metric, $J$-metric, ESP, $|\Delta N|$, DORI)} and after the \aposteriori refitting correction (\textit{DORI-refit}).
}}
\begin{ruledtabular}
\begin{tabular}{ccccccc}
\multirow{2}{*}{Model}   &$S$-metric               & $J$-metric           & ESP                   & $|\Delta N|$     & DORI                                      & DORI-refit        \\ \cline{2-2}\cline{3-3}\cline{4-4}\cline{5-5}\cline{6-7}
                         & mean error, bohr$^{-3}$ & mean error, kcal/mol & mean error, kcal/mol  & mean error       & \multicolumn{2}{c}{mean $L(0.1,0.75) \pm \mathrm{stdev}$, \%} \\ \hline
$J$-fitting              & \num{4.23e-3}           & \num{3.39e-02}       & 0.05                  & \num{7.70E-04}   &  $ 82.7 \pm 1.8 $           &                                 \\
$S$-fitting              & \num{4.15e-3}           & \num{4.82e-02}       & 1.08                  & \num{1.72E-02}   &  $ 81.1 \pm 1.5 $           &                                 \\
$\bf M_0$, $JJ$          & \num{4.40e-3}           & \num{3.68e-01}       & 0.89                  & \num{1.20E-02}   &  $ 81.8 \pm 2.2 $           & $ 81.2 \pm 3.2 $                \\
$\bf M_0$, $JS$          & \num{4.37e-3}           & \num{5.67e-01}       & 1.98                  & \num{3.02E-02}   &  $ 82.3 \pm 2.1 $           & $ 82.3 \pm 2.3 $                \\
$\bf M_0$, $SJ$          & \num{4.32e-3}           & \num{4.09e-01}       & 1.47                  & \num{2.24E-02}   &  $ 81.1 \pm 2.1 $           & $ 79.8 \pm 4.0 $                \\
$\bf M_0$, $SS$          & \num{4.29e-3}           & \num{6.34e-01}       & 2.30                  & \num{3.57E-02}   &  $ 81.4 \pm 2.0 $           & $ 81.4 \pm 2.2 $                \\
\end{tabular}
\end{ruledtabular}
\end{table}

\begin{table}[h]
\caption{\label{tab:fit}{
Mean absolute $E_O$ and relative $e_O$ errors,
computed on the training set, for the $S$- and $J$- fitting schemes
without and with constraint on the number of electrons.
}}
\begin{ruledtabular}
\begin{tabular}{lcccccc}
                           & $E_S$, bohr$^{-3}$ & $e_S$, ppm  &  $E_J$, kcal/mol & $e_J$, ppm      \\ \hline
$S$-fitting                & \num{4.2e-3}       & 14.9        &       0.07       & 0.14\phantom{0} \\
$S$-fitting (constrained)  & \num{4.2e-3}       & 14.9        &       0.05       & 0.094           \\
$J$-fitting                & \num{4.2e-3}       & 15.2        &       0.03       & 0.068           \\
$J$-fitting (constrained)  & \num{4.2e-3}       & 15.2        &       0.03       & 0.068           \\
\end{tabular}
\end{ruledtabular}
\end{table}

\clearpage

Numerical data for Fig.~2 are shown in Table~\ref{tab:espdip}.

\begin{table}[h]
\caption{\label{tab:espdip}{
\textbf{Numerical data for Fig.~2}:
Comparison of four combinations of metrics ($JJ$, $JS$, $SJ$, and $SS$),
two models ($\bf M_0$ and $\bf M_K$),
and three ways to correct for the number of electrons ($\vec O$, $\vec 1$, and $\bf sc$):
mean errors in the electrostatic potential on the isosurface $p_0 = 4$
and in the dipole moments.
}}
\begin{ruledtabular}
\begin{tabular}{ccccccc}
\multirow{2}{*}{Model} &\multicolumn{3}{c}{Electrostatic potential (kcal/mol)}&\multicolumn{3}{c}{Dipole moment (a.u.)}\\
& $\vec O$ & $\vec 1$ & $\bf sc$ & $\vec O$ & $\vec 1$ & $\bf sc$ \\ \hline
$\bf M_0$, $JJ$  &  \num{4.19E-01}  &  \num{4.28E-01}  &  \num{4.39E-01}  &  \num{1.11E-01}  &  \num{1.10E-01}  &  \num{1.14E-01} \\
$\bf M_0$, $JS$  &  \num{4.86E-01}  &  \num{4.83E-01}  &  \num{5.16E-01}  &  \num{1.27E-01}  &  \num{1.25E-01}  &  \num{1.36E-01} \\
$\bf M_0$, $SJ$  &  \num{4.05E-01}  &  \num{4.10E-01}  &  \num{4.28E-01}  &  \num{9.57E-02}  &  \num{9.42E-02}  &  \num{1.02E-01} \\
$\bf M_0$, $SS$  &  \num{4.79E-01}  &  \num{4.79E-01}  &  \num{5.13E-01}  &  \num{1.18E-01}  &  \num{1.15E-01}  &  \num{1.29E-01} \\
$\bf M_K$, $JJ$  &  \num{2.53E-01}  &  \num{2.91E-01}  &  \num{3.22E-01}  &  \num{5.89E-02}  &  \num{5.86E-02}  &  \num{7.07E-02} \\
$\bf M_K$, $SS$  &  \num{4.06E-01}  &  \num{4.34E-01}  &  \num{4.67E-01}  &  \num{9.31E-02}  &  \num{1.00E-01}  &  \num{1.11E-01} \\
\end{tabular}
\end{ruledtabular}
\end{table}

\begin{table}[h]
\caption{\label{tab:espfull}{
Comparison of four combinations of metrics ($JJ$, $JS$, $SJ$, and $SS$),
two models ($\bf M_0$ and $\bf M_K$),
and three ways to correct for the number of electrons ($\vec O$, $\vec 1$, and $\bf sc$):
mean errors in the electrostatic potential on the isosurfaces with different $p_0$.
}}
\begin{ruledtabular}
\begin{tabular}{ccccccc}
Model                         &   $p_0 = 1/16$      &    $p_0 = 1/4$        &   $p_0 = 1$         &    $p_0 = 4$         & $p_0 = 32$        \\ \hline
$\bf M_0$, $SS$  ($\vec  O$)  &   \num{2.58E-01}    &    \num{3.04E-01}     &   \num{3.70E-01}    &    \num{4.79E-01}    &    \num{9.08E-01} \\
$\bf M_0$, $SJ$  ($\vec  O$)  &   \num{2.15E-01}    &    \num{2.54E-01}     &   \num{3.11E-01}    &    \num{4.05E-01}    &    \num{7.72E-01} \\
$\bf M_0$, $JS$  ($\vec  O$)  &   \num{2.68E-01}    &    \num{3.13E-01}     &   \num{3.80E-01}    &    \num{4.86E-01}    &    \num{9.08E-01} \\
$\bf M_0$, $JJ$  ($\vec  O$)  &   \num{2.32E-01}    &    \num{2.71E-01}     &   \num{3.28E-01}    &    \num{4.19E-01}    &    \num{7.82E-01} \\
$\bf M_K$, $SS$  ($\vec  O$)  &   \num{2.13E-01}    &    \num{2.52E-01}     &   \num{3.10E-01}    &    \num{4.06E-01}    &    \num{7.86E-01} \\
$\bf M_K$, $JJ$  ($\vec  O$)  &   \num{1.33E-01}    &    \num{1.57E-01}     &   \num{1.93E-01}    &    \num{2.53E-01}    &    \num{5.23E-01} \\
$\bf M_0$, $SS$  ($\vec  1$)  &   \num{2.56E-01}    &    \num{3.02E-01}     &   \num{3.69E-01}    &    \num{4.79E-01}    &    \num{9.15E-01} \\
$\bf M_0$, $SJ$  ($\vec  1$)  &   \num{2.16E-01}    &    \num{2.55E-01}     &   \num{3.14E-01}    &    \num{4.10E-01}    &    \num{7.95E-01} \\
$\bf M_0$, $JS$  ($\vec  1$)  &   \num{2.65E-01}    &    \num{3.10E-01}     &   \num{3.76E-01}    &    \num{4.83E-01}    &    \num{9.09E-01} \\
$\bf M_0$, $JJ$  ($\vec  1$)  &   \num{2.34E-01}    &    \num{2.74E-01}     &   \num{3.33E-01}    &    \num{4.28E-01}    &    \num{8.05E-01} \\
$\bf M_K$, $SS$  ($\vec  1$)  &   \num{2.28E-01}    &    \num{2.70E-01}     &   \num{3.31E-01}    &    \num{4.34E-01}    &    \num{8.62E-01} \\
$\bf M_K$, $JJ$  ($\vec  1$)  &   \num{1.48E-01}    &    \num{1.76E-01}     &   \num{2.19E-01}    &    \num{2.91E-01}    &    \num{6.14E-01} \\
$\bf M_0$, $SS$  ($\bf  sc$)  &   \num{2.78E-01}    &    \num{3.27E-01}     &   \num{3.97E-01}    &    \num{5.13E-01}    &    \num{9.68E-01} \\
$\bf M_0$, $SJ$  ($\bf  sc$)  &   \num{2.28E-01}    &    \num{2.69E-01}     &   \num{3.29E-01}    &    \num{4.28E-01}    &    \num{8.22E-01} \\
$\bf M_0$, $JS$  ($\bf  sc$)  &   \num{2.85E-01}    &    \num{3.33E-01}     &   \num{4.03E-01}    &    \num{5.16E-01}    &    \num{9.60E-01} \\
$\bf M_0$, $JJ$  ($\bf  sc$)  &   \num{2.41E-01}    &    \num{2.83E-01}     &   \num{3.42E-01}    &    \num{4.39E-01}    &    \num{8.24E-01} \\
$\bf M_K$, $SS$  ($\bf  sc$)  &   \num{2.48E-01}    &    \num{2.92E-01}     &   \num{3.58E-01}    &    \num{4.67E-01}    &    \num{9.27E-01} \\
$\bf M_K$, $JJ$  ($\bf  sc$)  &   \num{1.67E-01}    &    \num{1.98E-01}     &   \num{2.45E-01}    &    \num{3.22E-01}    &    \num{6.65E-01} \\
\end{tabular}
\end{ruledtabular}
\end{table}

\clearpage

However, the error values have a wide spread (mean $\approx$ standard deviation),
and do not obey the normal distribution.
Even though Fig.~2 shows the mean values as an illustration, additional analysis is needed to
show the statistical relevance of the error differences.
Table~\ref{tab:wilc} shows the $p$-values obtained with the Wilcoxon matched-pairs signed rank test\cite{W1945}
computed using the \texttt{R}\cite{R2018} \texttt{stats} module implementation.\cite{B1972,HW1973}

\begin{table}[h]
\caption{\label{tab:wilc}{
$p$-values obtained with Wilcoxon test for the test set errors in the ESP and dipole moments
within the $\bf M_0$ and $\bf M_K$ models (refitting correction for $N$) and all the metric combinations used in the work.
The cells are colored in \colorbox{mytablegreen}{green} when $p<0.05$ and in \colorbox{mytablered}{red} when $p>0.95$.
}}
\begin{ruledtabular}
\begin{tabular}{ccccccccccccccc}
\multirow{2}{*}{Model}
               && \multicolumn{6}{c}{ESP}                                                                                  && \multicolumn{6}{c}{dipole moments}                                                                        \\ \cline{3-8}\cline{10-15}
               && $\bf M_0$, $SS$ & $\bf M_0$, $SJ$ & $\bf M_0$, $JS$ & $\bf M_0$, $JJ$ & $\bf M_K$,$SS$ & $\bf M_K$, $JJ$ && $\bf M_0$, $SS$ & $\bf M_0$, $SJ$ & $\bf M_0$, $JS$ & $\bf M_0$, $JJ$ & $\bf M_K$, $SS$ & $\bf M_K$, $JJ$ \\ \hline
$\bf M_0$,$SS$ && $\times$        & \zz{0.00}       & \zz{0.18}       & \zz{0.00}       & \zz{0.00}      & \zz{0.00}       && $\times$        & \zz{0.00}       & \zz{1.00}       & \zz{0.00}       & \zz{0.00}       & \zz{0.00}       \\
$\bf M_0$,$SJ$ && \zz{1.00}       & $\times$        & \zz{1.00}       & \zz{0.02}       & \zz{0.51}      & \zz{0.00}       && \zz{1.00}       & $\times$        & \zz{1.00}       & \zz{1.00}       & \zz{0.85}       & \zz{0.00}       \\
$\bf M_0$,$JS$ && \zz{0.82}       & \zz{0.00}       & $\times$        & \zz{0.00}       & \zz{0.00}      & \zz{0.00}       && \zz{0.00}       & \zz{0.00}       & $\times$        & \zz{0.00}       & \zz{0.00}       & \zz{0.00}       \\
$\bf M_0$,$JJ$ && \zz{1.00}       & \zz{0.98}       & \zz{1.00}       & $\times$        & \zz{0.98}      & \zz{0.00}       && \zz{1.00}       & \zz{0.00}       & \zz{1.00}       & $\times$        & \zz{0.04}       & \zz{0.00}       \\
$\bf M_K$,$SS$ && \zz{1.00}       & \zz{0.49}       & \zz{1.00}       & \zz{0.02}       & $\times$       & \zz{0.00}       && \zz{1.00}       & \zz{0.15}       & \zz{1.00}       & \zz{0.96}       & $\times$        & \zz{0.00}       \\
$\bf M_K$,$JJ$ && \zz{1.00}       & \zz{1.00}       & \zz{1.00}       & \zz{1.00}       & \zz{1.00}      & $\times$        && \zz{1.00}       & \zz{1.00}       & \zz{1.00}       & \zz{1.00}       & \zz{1.00}       & $\times$        \\
\end{tabular}
\end{ruledtabular}
\end{table}

If the $p$-value is less than a significant level $\alpha$,
we can conclude that the median of the values of the first sample (column) is significantly smaller than the median of the values for the second sample (row);
if $p > 1-\alpha$, the first sample is significantly larger than the second sample.
With $\alpha = 0.05$, the ESP the errors are ranged as
\begin{equation}
{\bf MK}, JJ < {\bf M0}, JJ < {\bf M0}, SJ \approx {\bf MK}, SS < {\bf M0}, JS \approx {\bf M0}, SS
\end{equation}
and for the dipole moments as
\begin{equation}
{\bf MK}, JJ < {\bf M0}, SJ \approx {\bf MK}, SS < {\bf M0}, JJ < {\bf M0}, SS < {\bf M0}, JS
\end{equation}
and ($\bf MK$, $JJ$) is the best model tested.

\clearpage
\section{Extrapolation}

\begin{figure*}[h]
\centering
\includegraphics[width=0.98\textwidth]{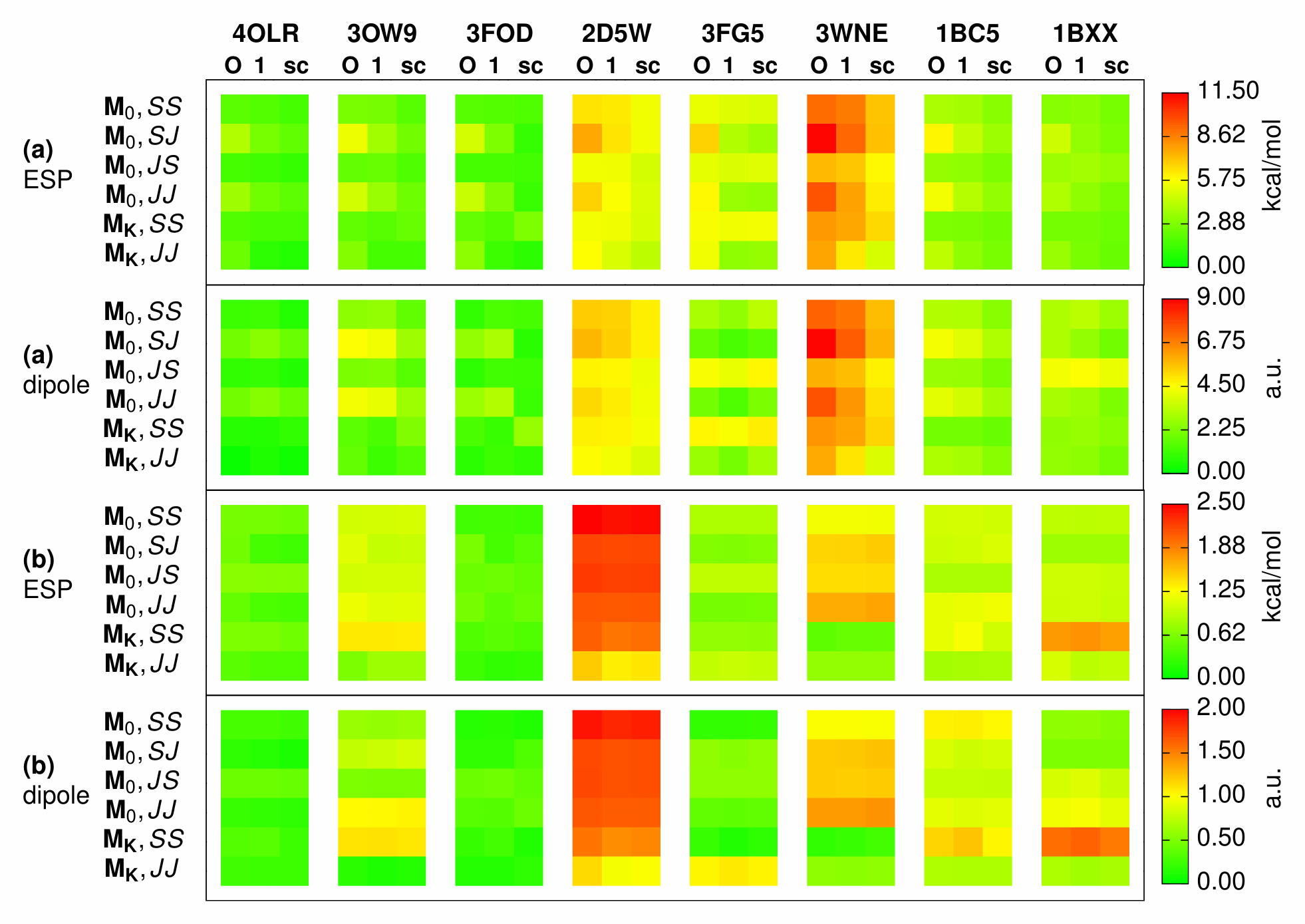}
\caption{
\textbf{Additional data for Fig.~3}:
Comparison of the four combinations of metrics ($JJ$, $JS$, $SJ$, and $SS$),
two models ($\bf M_0$ and $\bf M_K$),
and three ways to correct for the number of electrons ($\vec O$, $\vec 1$, and \textbf{sc})
by prediction for eight oligopeptides (labeled by PDB~ID).
Each square represents an error in predicted
electrostatic potential on the isosurface $p_0 = 0.125$ or
dipole moments of \textit{(a)} oligopeptides and \textit{(b)} ``no-backbone'' oligopeptides.
Unlike Fig.~3, here the absolute errors are presented.
}
\label{fig:oligo}
\end{figure*}

Both 2D5W and 3WNE contain the proline residue (see Table~\ref{tab:oligo} for the composition of the oligopeptides),
which is not represented in the training set and can be significantly distorted after cutting. In addition, 3WNE is the only cyclic structure on which a prediction is made.
Its strained backbone has thus not been seen by the model. This can explain the high absolute errors (Fig.~\ref{fig:oligo}) for these two oligopeptides.

\begin{table}[h]
\caption{\label{tab:oligo}{Composition of the oligopeptides}}
\begin{minipage}{0.5\linewidth}
\begin{ruledtabular}
\begin{tabular}{lll}
\multicolumn{1}{c}{PDB ID} & \multicolumn{2}{c}{Sequence}        \\ \hline
4OLR (4OLR\_1)             &  Tyr Val Val Phe Val     & YVVFV   \\
3OW9 (3OW9\_1)             &  Lys Leu Val Phe Phe Ala & KLVFFA  \\
3FOD (3FOD\_1)             &  Ala Ile Leu Ser Ser Thr & AILSST  \\
2D5W (2D5W\_2)             &  Ala Ser Lys Pro Lys     & ASKPK   \\
3FG5 (3FG5\_2)             &  Phe Leu Ser Tyr Lys     & FLSYK   \\
3WNE (3WNE\_2)             &  Pro Lys Ile Asp Asn Gly & PKIDNG  \\
1BC5 (1BC5\_2)             &  Asn Trp Glu Thr Phe     & XNWETF  \\
1BXX (1BXX\_2)             &  Asp Tyr Gln Arg Leu Asn & DYQRLN  \\
\end{tabular}
\end{ruledtabular}
\end{minipage}
\end{table}

\begin{table}[p]
\caption{\label{tab:oligoN}{
The errors in the number of electrons $\Delta N$ for the predictions of the original model $\bf M_0$ before the \aposteriori correction
and the mean absolute errors (MAE) for each metric set
along with the total number of electrons $N$,
both for the original oligopeptides and the ``no-backbone'' structures.
}}
\begin{ruledtabular}
\begin{tabular}{lldddddddddddd}
           &                   & \multicolumn{8}{c}{\it Original oligopeptides}                                                   \\
           &                   & \bf 4OLR  & \bf 3OW9 & \bf 3FOD & \bf 2D5W & \bf 3FG5 & \bf 3WNE & \bf 1BC5 & \bf 1BXX & \rm MAE \\ \hline
\multicolumn{2}{c}{Total $N$}  & 336       &  390     & 318      &  286     &  352     &  334     &  368     &  430     &         \\
\multirow{4}{*}{$\Delta N$}
           &  $\bf M_0$, $SS$  & 1.12      &  1.56    & 1.50     &  1.40    &  0.90    &  1.83    &  1.02    &  1.21    & 1.30    \\
           &  $\bf M_0$, $SJ$  & 1.18      &  1.50    & 1.43     &  1.46    &  1.23    &  2.10    &  1.28    &  1.42    & 1.03    \\
           &  $\bf M_0$, $JS$  & 0.83      &  1.21    & 1.21     &  1.11    &  0.62    &  1.32    &  0.83    &  0.87    & 1.49    \\
           &  $\bf M_0$, $JJ$  & 1.05      &  1.33    & 1.31     &  1.28    &  1.06    &  1.73    &  1.18    &  1.21    & 1.35    \\ \hline\hline
                                                                                                                        &         \\ \hline\hline
           &                  & \multicolumn{8}{c}{\it No-backbone oligopeptides}                                       &         \\
           &                  & \bf 4OLR  & \bf 3OW9 & \bf 3FOD & \bf 2D5W & \bf 3FG5 & \bf 3WNE & \bf 1BC5 & \bf 1BXX  & \rm MAE \\ \hline
\multicolumn{2}{c}{Total $N$} & 226       &  260     & 188      &  176     &  242     &  212     &  258     &  300      &         \\
\multirow{4}{*}{$\Delta N$}
           & $\bf M_0$, $SS$  & -0.10     &  0.11    & 0.06     &  0.39    &  -0.01   &  0.02    &  -0.11   &  0.13     & 0.07    \\
           & $\bf M_0$, $SJ$  & -0.13     &  -0.11   & -0.13    &  0.11    &  -0.06   &  -0.10   &  -0.14   &  -0.01    & 0.11    \\
           & $\bf M_0$, $JS$  & -0.09     &  0.09    & 0.07     &  0.34    &  -0.02   &  -0.03   &  -0.08   &  0.13     & 0.10    \\
           & $\bf M_0$, $JJ$  & -0.08     &  -0.06   & -0.09    &  0.12    &  -0.02   &  -0.08   &  -0.07   &  0.06     & 0.12    \\
\end{tabular}
\end{ruledtabular}
\end{table}

\begin{figure*}[p]
\centering
\includegraphics[width=0.666 \textwidth]{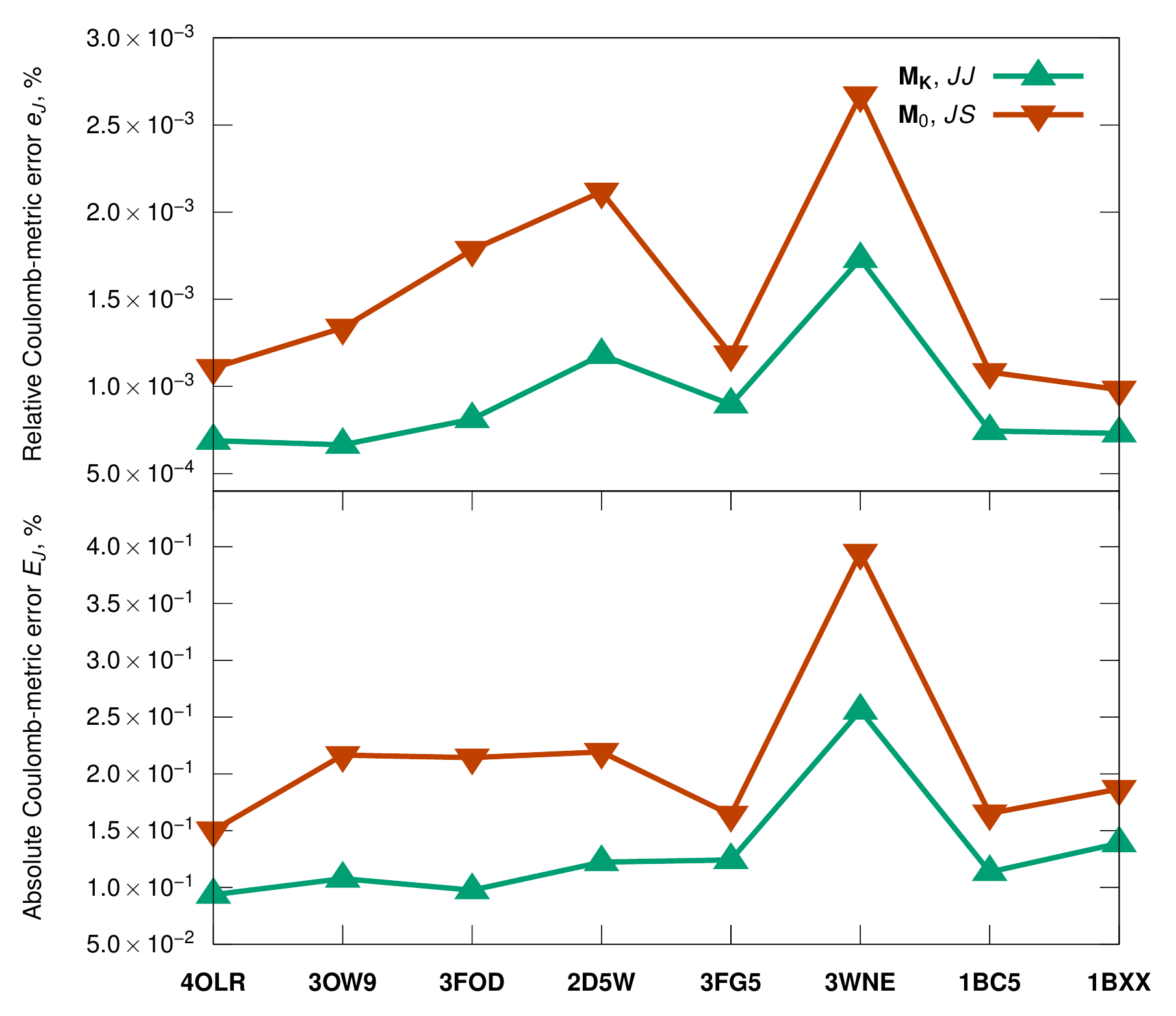}
\caption{
Comparison of the Coulomb-metric errors ($e_J$ and $E_J$)
between the \abinitio and the predicted densities using the original not-corrected $\bf M_0$ model ($JS$-scheme)
and the corrected $\bf M_K$ model ($JJ$-scheme) for the oligopepdides in Ref.~\onlinecite{FGMCC2019}.
Except for the more challenging cyclopeptide
(3WNE),\cite{FGMCC2019} the varying trends essentially arise from
the different protein sizes (and number of electrons) (see Table~\ref{tab:oligoN}).
}
\label{fig:pept}
\end{figure*}

\clearpage
\section{Comparison of \texorpdfstring{$L^2$}{L2} and \texorpdfstring{$L^1$}{L1} norms}

\begin{figure*}[h]
\centering
\includegraphics[width=0.666 \textwidth]{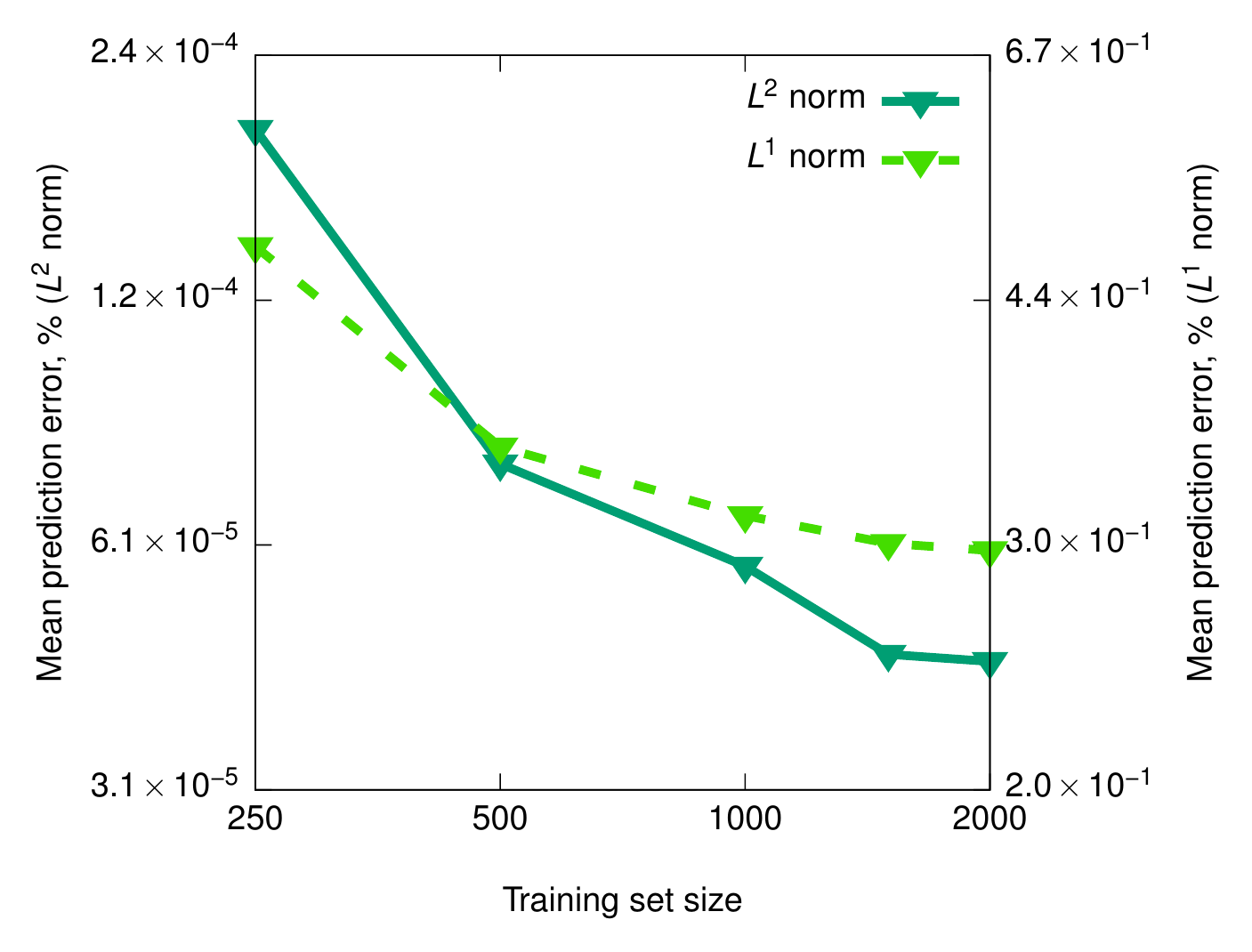}
\caption{
Comparison of the learning curves computed using
the $L^2$ overlap norm (dark green, solid line, left $y$-axis)
and the $L^1$ overlap norm (light green, dashed line, right $y$-axis)
on the example of the $JS$ metric combination and $\bf M_0$ model.
The number of reference environments $M=1000$.
}
\label{fig:lcL1}
\end{figure*}

The use of the $L^2$ norm (Eq. \ref{eq:l2}) to define the error in the main text arises naturally
from the quadratic form of both the decomposition and the learning loss functions.
Nonetheless, in the past,\cite{FGMCC2019} we have explored the possibility to report the density error
using the $L^1$ norm (Eq. \ref{eq:l1}), which has the advantage to be more robust
with respect to the presence of eventual outliers in the dataset.
To establish a more clear connection between the results presented in this work and our previous research,
we show in Fig.~\ref{fig:lcL1} the numerical value of the $L^1$ and $L^2$ errors for the original $JS$ model
used for instance in Ref.~\citenum{FGMCC2019}.
Not surprisingly, the quadratic form of the $L^2$ norm leads to smaller absolute values of the error and a steeper curve,
due to the increased sensitivity to density variability in the dataset.

\begin{equation}
\label{eq:l2}
L^2\mathrm{error}
= e_S = \frac{\displaystyle\int\big(\rho_\mathrm{ML}(\vec r)-\rho_\mathrm{DF}(\vec r)\big)^2 \de^3\vec r}
{\displaystyle\int \rho_\mathrm{DF}^2(\vec r) \de^3\vec r}
\end{equation}

\begin{equation}
\label{eq:l1}
L^1\mathrm{error}
= \frac{\displaystyle\int\big|\rho_\mathrm{ML}(\vec r)-\rho_\mathrm{DF}(\vec r)\big| \de^3\vec r}
{\displaystyle\int \big|\rho_\mathrm{DF}(\vec r)\big| \de^3\vec r}
\end{equation}

\clearpage
\section{Performance of \texorpdfstring{$\bf M_0$}{M\_0} on dipole moments}

Since the original ML model ($\bf M_0$)\cite{GFMWCC2019,FGMCC2019} is not constrained to predict densities
with the exact number of electrons, they do not integrate to exactly cancel out the sum of nuclear charges,
leaving the molecules partially charged. For charged systems, dipole moments and higher multipoles are not uniquely defined
and their values depends on the origin of the chosen frame of reference.
In practice, this means that it is possible to obtain any value of the dipoles (including the correct one)
just by choosing an \textit{ad hoc} origin of the coordinate system for each molecule.
As the choice of origin is entirely arbitrary, we do not use dipole moments to assess the effects of the metric in the main text.

For completeness, however, we report here the performance of $\bf M_0$ for the prediction of dipole moments
with all possible metric combinations by fixing the origin of the coordinate system to the nuclear center of charge.
Nonetheless, since this coordinate choice is as good as any other, we do not discuss further the performance of each metric combination.

\begin{figure*}[h]
\centering
\includegraphics[width=0.33\textwidth]{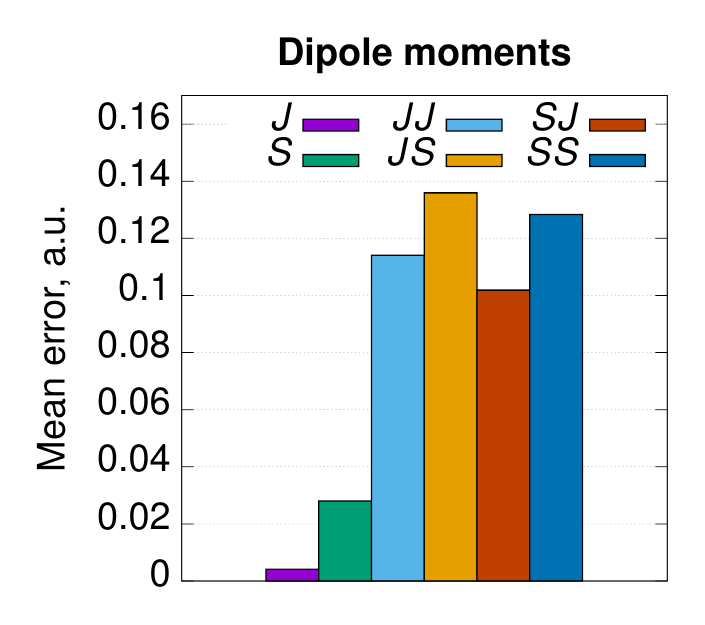}
\caption{
Mean errors in dipole moments computed on the test set for fitted ($J$ and $S$) and predicted ($JJ$, $JS$, $SJ$, and $SS$) densities
within the $\bf M_0$ model without any constraints on the number of electrons,
computed with respect to the corresponding $\rho_\mathrm{QM}$.
}
\label{fig:diplole}
\end{figure*}

\section*{References}
\bibliography{sj}